\documentclass[journal]{IEEEtran}
\IEEEoverridecommandlockouts

\usepackage{color}
\usepackage{bm}
\usepackage{graphicx}
\usepackage{amsmath}
\usepackage{amssymb}
\usepackage[ruled]{algorithm2e}
\usepackage{multirow}
\usepackage{booktabs}
\usepackage{array}
\usepackage{lipsum}
\usepackage{enumerate}
\usepackage{stfloats}
\usepackage{subfigure}
\usepackage{cases}
\usepackage{diagbox}
\usepackage{cite}

\usepackage{enumitem}
\usepackage {verbatim}

\usepackage{mathrsfs}

\SetKwRepeat{APDo}{AP side ($a \in \mathcal{A}),$ $\backslash\backslash\mathtt{In\;Parallel}$}{end}
\SetKwRepeat{CPUDo}{CPU side}{end}

\newtheorem{remark}{Remark}
\newtheorem{prop}{Proposition}

\newtheorem{theorem}{Theorem}
\newtheorem{lemma}{Lemma}
\newtheorem{corollary}{Corollary}

\allowdisplaybreaks[4]

\begin{document}

\title{Cooperative Integrated Sensing and Communication Networks: Analysis and Distributed Design}

\author{Bowen Wang, \IEEEmembership{Student Member, IEEE},
		Hongyu Li, \IEEEmembership{Member, IEEE},
		Fan Liu, \IEEEmembership{Senior Member, IEEE}, \\
        Ziyang Cheng,~\IEEEmembership{Senior Member, IEEE}
        and Shanpu Shen, \IEEEmembership{Senior Member, IEEE}.
        \vspace{-2em}
\thanks{B. Wang and Z. Cheng are with the School of Information and Communication Engineering, University of Electronic Science and Technology of China, Chengdu 611731, China. (e-mail: bwwang@ieee.org, zycheng@uestc.edu.cn).}
\thanks{H. Li is with the Internet of Things Thrust, The Hong Kong University of Science and Technology (Guangzhou), Guangzhou 511400, China (e-mail: hongyuli@hkust-gz.edu.cn).}
\thanks{F. Liu is with the National Mobile Communications Research Laboratory, School of Information Science and Engineering, Southeast University, Nanjing 210096, China (e-mail: fan.liu@seu.edu.cn).}
\thanks{S. Shen is with the Department of Electrical Engineering \& Electronics, University of Liverpool, Liverpool L69 3GJ, U.K. (email: Shanpu.Shen@liverpool.ac.uk).}
}

\def\refmylabel{3}

\maketitle

\begin{abstract}
	This paper proposes a cooperative integrated sensing and communication network (Co-ISACNet) adopting hybrid beamforming (HBF) architecture, which improves both radar sensing and communication performance. 
	The main contributions of this work are four-fold. 
	First, we introduce a novel cooperative sensing method for the considered Co-ISACNet, followed by a comprehensive analysis of this method. 
	This analysis mathematically verifies the benefits of Co-ISACNet and provides insightful design guidelines.
	Second, to show the benefits of Co-ISACNet, we propose to jointly design the HBF to maximize the network communication capacity while satisfying the constraint of beampattern similarity for radar sensing, which results in a highly dimensional and non-convex problem.
	Third, to facilitate the joint design, we propose a novel distributed optimization framework based on proximal gradient and alternating direction method of multipliers, namely PANDA.
	Fourth, we further adopt the proposed PANDA framework to solve the joint HBF design problem for the Co-ISACNet.
	By using the proposed PANDA framework, all access points (APs) optimize the HBF in parallel, where each AP only requires local channel state information and limited message exchange among the APs.
    Such framework reduces significantly the computational complexity and thus has pronounced benefits in practical scenarios.
	Simulation results verify the effectiveness of the proposed algorithm compared with the conventional centralized algorithm and show the remarkable performance improvement of radar sensing and communication by deploying Co-ISACNet.
\end{abstract}


 \vspace{-1em}

\section{Introduction}

\IEEEPARstart{T}{he}
explosive growth of wireless services and the severe spectrum shortage have driven the demand for new paradigms and technologies to overcome spectrum congestion and improve spectrum efficiency for future wireless networks \cite{Liu2022Overview}.
Among all of the emerging techniques, integrated sensing and communications (ISAC), where the radar sensing and wireless communication operations are integrated and jointly designed in a common hardware platform \cite{cui2021integrating}, has benefits in enhancing spectrum efficiency and reducing hardware cost. 
With these benefits, ISAC is promising in supporting industry 4.0, autonomous vehicles and Internet-of-Things (IoT) in future wireless networks \cite{zhang2021overview}.

Extensive research results have focused on the waveform design for ISAC systems in sub-6GHz frequencies, where the transmitter is employed with fully-digital beamforming architectures \cite{liu2018toward,liu2020joint,liu2021cramer,xu2021rate,chen2021joint,wang2018dual}. 
To achieve higher-precision sensing while guaranteeing higher-throughput wireless communications, the research for ISAC has moved to millimeter wave (mmWave) frequency. 
The shorter wavelength of mmWave signals together with massive multiple input multiple output (MIMO)  may provide sufficient gains to combat the severe path loss, while the fully-digital beamforming architecture in sub-6GHz bands is not viable for mmWave frequencies due to the extensively increasing cost and power consumption of radio frequency (RF) chains and other hardware components \cite{heath2016overview,ahmed2018survey,cheng2022hybrid}. 
To address the above issue, hybrid beamforming (HBF) architecture \cite{heath2016overview,ahmed2018survey,cheng2022hybrid}, which partitions the beamforming operation into a small-dimensional digital beamforming and a large-dimensional analog beamforming realized by a phase shift network, is proposed to compensate for the severe path loss with affordable cost and power consumption \cite{liu2020jointTCOM,liu2020jointTCOM,cheng2022double,wang2022exploiting,zeng2022joint,Barneto2022TCOM,Islam2022Simultaneous,Qi2022COML}.
Prior work on HBF design for mmWave ISAC system is carried out in \cite{liu2020jointTCOM}, where a novel transceiver with HBF architecture for a ISAC base station (BS) is proposed.
Following \cite{liu2020jointTCOM}, ISAC with double-phase-shifter-based HBF architecture is investigated in \cite{cheng2022double}, where the target detection performance of the extended target is improved while ensuring the downlink communication performance.
In addition, a symbol-level based HBF design is proposed in \cite{wang2022exploiting}, where the constructive interference is utilized to improve both sensing and communication performance.
To further improve the communication rate, the authors in \cite{zeng2022joint,Barneto2022TCOM,Islam2022Simultaneous} investigate the HBF design for a wideband OFDM ISAC system.

Nevertheless, the above-mentioned mmWave ISAC works \cite{liu2020jointTCOM,cheng2022double,wang2022exploiting,zeng2022joint} are restricted to single-source scenarios, limiting the coverage for both wireless sensing and communications. 
Fortunately, there are already solutions in mmWave communications, that is to establish ultra-dense networks, namely cooperative cell-free (CoCF) networks where all access points (APs) controlled by a central processing unit (CPU) cooperatively serve users without cell boundaries \cite{lee2016hybrid,wang2022joint,al2022self,zhang2019cell,bjornson2013optimal,li2014multi,Jafri2022TCOM}, thereby offering seamless wireless coverage.
Relative works in mmWave CoCF \cite{lee2016hybrid,wang2022joint,al2022self,zhang2019cell,bjornson2013optimal,li2014multi,Jafri2022TCOM} have demonstrated the benefits of improving wireless communication quality and enlarging coverage.
The constructive results in mmWave CoCF \cite{lee2016hybrid,wang2022joint,al2022self,zhang2019cell,bjornson2013optimal,li2014multi,Jafri2022TCOM} motivate us to employ CoCF networks in mmWave/THz ISAC, namely cooperative ISAC network (Co-ISACNet) \cite{wang2020constrained,chen2023joint,Cheng2023Optimal}, which has benefits from the following perspectives:
1) From the communication perspective, the CoCF network enables cooperative signal transmission \cite{zhang2019cell,bjornson2013optimal,li2014multi}, thus enabling high-quality mmWave communications.
2) From the radar perspective, multiple distributed APs in CoCF network provide extra spatial degrees-of-freedom (DoFs) \cite{haimovich2007mimo,shen2010fundamental,win2011network,Rahman2020}, which potentially leads to improved sensing performance.

Despite the above benefits, the mmWave Co-ISACNet encounters several challenges:
1) The system model and operational mechanism for mmWave Co-ISACNet are still underdeveloped. 
While CoCF has been proved to have benefits in wireless communication networks, synergizing sensing functions with CoCF networks to achieve Co-ISACNet remains a challenging open problem.
This raises new technical issues, e.g., how to effectively leverage the capabilities of multiple APs to achieve \textit{cooperative sensing}.
2) The beamforming design inevitably faces challenges due to the increasing number of sources and antennas, complex constraints, and multiple functionalities. In this sense, the conventional \textit{centralized design} for multi-source networks, where the beamforming design for different sources is centralized at the CPU \cite{chang2016asynchronous,chang2014multi,shi2015extra,shi2015proximal}, would result in heavy computational burden and thus is not practical for the mmWave Co-ISACNet.

In this paper, we aim to address the above challenges to demonstrate the benefits of the mmWave Co-ISACNet.
To this end, we introduce a novel \textit{cooperative sensing} method.
Additionally, we propose a novel \textit{distributed optimization} algorithm to jointly design the HBF for the mmWave Co-ISACNet, which has affordable beamforming design complexity with a reduced computational cost at the CPU.
Our contributions are summarized as follows:

\textit{First,} we propose a mmWave Co-ISACNet with HBF architecture, which consists of a CPU and multiple dual-function APs.
Then, we, for the first time, propose a practical \textit{cooperative sensing} method for Co-ISACNet.
Through comprehensive analysis, we reveal the essence of the proposed \textit{cooperative sensing} method, and shed light on valuable design insights.

\textit{Second,} to show the advantages of Co-ISACNet and validate our proposed cooperative sensing method, a joint HBF design problem is formulated to maximize the average sum rate for the considered Co-ISACNet, while satisfying the constraint of beampattern similarity for radar sensing.

\textit{Third,} we, for the first time, propose a general \textit{distributed optimization} algorithm, namely PANDA framework, to jointly design the HBF.
Particularly, the PANDA modifies the centralized alternating direction method of multipliers (ADMM) by the proximal gradient (PG) to decouple variables and enable distributed optimization.
Moreover, we theoretically prove that the proposed PANDA shares the same convergence behaviors as the conventional centralized ADMM.

\textit{Fourth,} we customize the proposed PANDA framework to solve the joint HBF design problem of the Co-ISACNet.
Specifically, we first reformulate the original problem into a tractable form by fractional programming.
Then, we apply the PANDA framework to tackle the reformulated problem, which requires only local CSI and minimal message exchange among the APs, thus reducing significantly the computational burden of the CPU as well as the backhaul signaling overheads.

\textit{Finally,} we present simulation results to evaluate the performance of the proposed algorithm and the Co-ISACNet.
Specifically, the proposed distributed HBF design algorithm can achieve nearly the same performance as the conventional centralized algorithm, which validates the efficiency of the proposed algorithm.
Furthermore, compared with the conventional ISAC with single AP, the proposed Co-ISACNet achieves better communication and radar performance, demonstrating the superiority of the proposed Co-ISACNet.

\textit{Organization}:
Section \ref{Sec-2} illustrates the system model of the proposed Co-ISACNet.
Section \ref{Sec-3} proposes a general distributed framework named PANDA.
Section \ref{Sec-4} adopts the proposed PANDA to solve the joint HBF design problem.
Section \ref{Sec-5} evaluates the performance of the proposed design and Section \ref{Sec-6} concludes this work.

\textit{Notations}:
Boldface lower- and upper-case letters indicate column vectors and matrices, respectively.
$\mathbb{C}$, $\mathbb{R}$, and $\mathbb{R}^+$ denote the set of complex numbers, real numbers, and positive real numbers, respectively.
$\mathbb{E}\{\cdot\}$ represents statistical expectation.
$(\cdot)^\ast$, $(\cdot)^T$, $(\cdot)^H$, and $(\cdot)^{-1}$ denote the conjugate, transpose, conjugate-transpose operations, and inversion, respectively.
$\Re \{ \cdot \}$ denotes the real part of a complex number.
$\mathbf{I}_L$ indicates an $L \times L$ identity matrix.
$\| \mathbf{A} \|_F$ denotes the Frobenius norm of matrix $\mathbf{A}$.
$|a|$ denotes the norm of variable $a$.
$\jmath = \sqrt{-1}$ denotes imaginary unit.
$\lambda_\mathrm{max} ( \mathbf{A} )$ is the maximum eigenvalue of $\mathbf{A}$.
$\angle \left( \mathbf{A} \right)$ denotes the phase values of $\mathbf{A}$.
$\otimes$ is the Kronecker product operator.
$\mathsf{diag}(\cdot)$ denotes a diagonal matrix.
$\mathsf{Tr}\{\cdot\}$ denotes the summation of diagonal elements of a matrix. 
Finally, $[\mathbf{A}]_{i,j}$, and $[\mathbf{a}]_i$ denote the $(i,j)$-th element of matrix $\mathbf{A}$, and the $i$-th element of vector $\mathbf{a}$, respectively.

\section{System Model and Problem Formulation}\label{Sec-2}

In this section, we describe operating mechanism of the Co-ISACNet, introduce the system model and derive performance metrics, and formulate the optimization problem.

\begin{figure}
	\centering
	\includegraphics[width=0.7\linewidth]{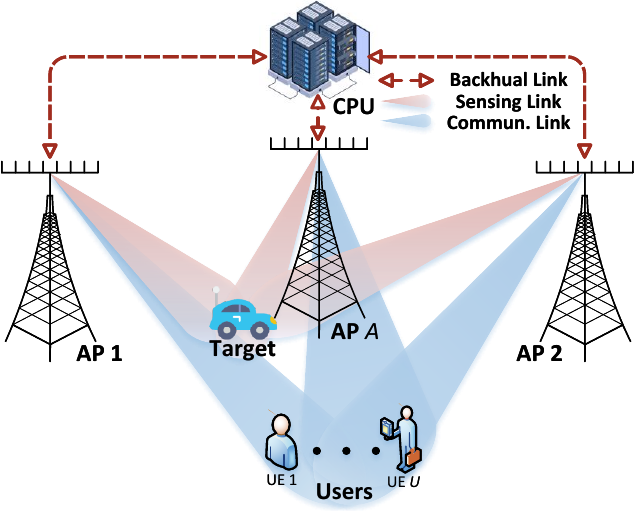}
	\vspace{-0.5em}
	\caption{Diagram of the proposed mmWave Co-ISACNet.}
	 \vspace{-1em}
	\label{fig:model}
	\vspace{-0.5em}
\end{figure}

\vspace{-1em}
\subsection{Operating Mechanism and Transmit Model}
As shown in Fig. \ref{fig:model}, we consider a Co-ISACNet comprising a CPU, a set of dual-function APs $\mathcal{A} = \{1,\cdots,A\}$, multiple downlink user equipments (UEs) $\mathcal{U} = \{1,\cdots,U\}$, radar targets $\mathcal{O} = \{ 1 , \cdots , O \}$ and clutter sources $\mathcal{Q} = \{1,\cdots,Q \}$.
Suppose that all APs are equipped with $N_\mathrm{T}$ transmit antennas and $N_\mathrm{R}$ receive antennas arranged as uniform linear arrays (ULA).
Multiple distributed APs cooperatively transmit waveforms to detect $O$ radar targets and simultaneously provide communication service to $U$ single-antenna downlink UEs.
The CPU is deployed for control and planning, to which all APs are connected by optical cables or wireless backhaul \cite{zhang2019cell,bjornson2013optimal,li2014multi}. 
The operation mechanism of the proposed scheme is concluded as the following three phases: 

\textit{Phase 1: Uplink Training.} In this phase, each UE is assigned a random pilot from a set of mutually orthogonal pilots utilized by the APs.
After correlating the received signal at AP $a$ ($a$-th AP), the estimate of channel ${\bf h}_{a,u} \in {\mathbb{C}}^{N_{\rm T}}$ from UE $u$ to AP $a$ can be performed by many existing methods \cite{zhang2019cell,bjornson2013optimal,li2014multi}, such as minimum mean square error (MMSE) estimator.

\textit{Phase 2: ISAC Transmission.} 
In this phase, according to the estimated CSI, APs first optimize the ISAC waveforms to maximize the network capacity and to ensure sensing performance.
Then, all APs transmit the ISAC signals towards the direction of UEs\footnote{In this paper, we assume that all APs are synchronized and scheduled to the same frame structure to achieve coherent downlink communication \cite{zhang2019cell,bjornson2013optimal,li2014multi}.} and targets.

\textit{Phase 3: Reception.} In this phase, the downlink UEs receive the signals from the APs and decode the received signals to obtain communication information.
The radar sensing receiver at each AP collects echo signals reflected by the targets to perform radar detection and estimation.

This paper focuses on the ISAC transmission and reception phases. 
In the following, we will elaborate on the ISAC transmitting model, derive ISAC performance metrics, and formulate the optimization problem.

\vspace{-1em}
\subsection{System Model}

\begin{figure}
	\centering
	\includegraphics[width=0.7\linewidth]{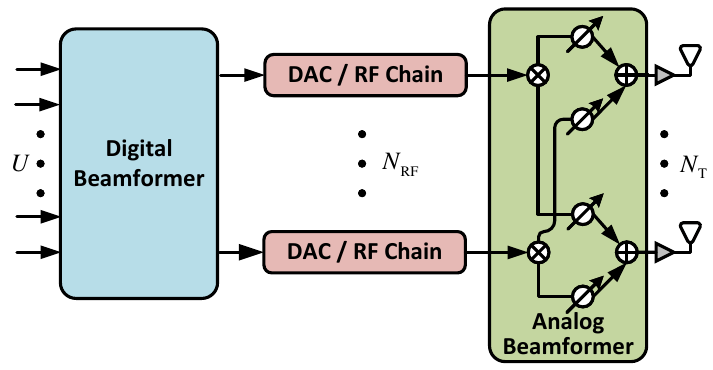}
	\vspace{-0.5em}
	\caption{HBF architecture with fully-connected phase shift network.}
	\vspace{-1em}
	\label{fig:HBF}
	\vspace{-0.5em}
\end{figure}

\subsubsection{Transmit Model}
We assume all APs are synchronized and serve multiple UEs by joint transmission.
Moreover, as shown in Fig. \ref{fig:HBF}, the APs are assumed to employ the HBF architecture with $N_{\rm RF}$ RF chains, $U \le N_\mathrm{RF} \ll N_\mathrm{T}$.
The HBF architecture consists of the digital beamformer $\mathbf{F}_{\mathrm{D},a} = [ \mathbf{f}_{\mathrm{D},a,1} , \cdots , \mathbf{f}_{\mathrm{D},a,U} ] \in {\mathbb{C}}^{N_\mathrm{RF} \times U}$ and the analog RF beamformer $\mathbf{F}_{\mathrm{A}, a} \in {\mathbb{C}}^{N_{\rm T} \times N_\mathrm{RF}}$ realized by a fully-connected phase shift network \cite{heath2016overview,ahmed2018survey,cheng2022hybrid} and thus, imposes a constant modulus constraint of each entry, i.e., $| {{[ {{\mathbf{F}_{\mathrm{A},a}}} ]}_{m,n}} | = 1, \forall m,n$.
Therefore, the transmitted signal from AP $a$ at time instant $t$ is
\begin{equation}
	\mathbf{x}_a(t) =  {\mathbf{F}_{\mathrm{A},a}} {\mathbf{F}_{\mathrm{D},a}} \mathbf{s}_{a}(t) = {\mathbf{F}_{\mathrm{A},a}} \sum\limits_{u \in \mathcal{U}}  {{\mathbf{f}_{\mathrm{D},a,u}}{s_{a,u}}(t)}    ,
	\label{eq:1}
\end{equation}
where $\mathbf{s}_{a}(t) = \sum_{l\in\mathcal{L}}{ \mathbf{s}_{a}[l]\mathrm{rect}\left(t-(l-1)\Delta t\right) }$ and ${s_{a,u}}(t) = \sum_{l\in\mathcal{L}}{ {s_{a,u}}[l]\mathrm{rect}\left(t-(l-1)\Delta t\right) } $ with $s_{a,u}(t)$ being the transmitted symbol to UE $u$ and $\mathbf{s}_a[l] = \left[s_{a,1}[l] , \cdots , s_{a,U}[l]\right]^T \in {\mathbb{C}}^U$ being the transmit symbols vector.
Since $\mathbf{s}_a[l]$ includes random communication symbols, we assume $\mathbb{E} \{\mathbf{s}_{a}[l_1]\mathbf{s}_{b}^H[l_2]\} = \mathbf{I}_{U}$ when $a=b$ and $l_1=l_2$, otherwise $\mathbb{E} \{\mathbf{s}_{a}[l_1]\mathbf{s}_{b}^H[l_2]\} = \mathbf{0}_{U}$.

\subsubsection{Communication Reception Model}

For the downlink communication, the received signal at UE $u$ is given by
\begin{equation}
	y_u(t) = \sum_{a\in \mathcal{A}}{\mathbf{h}_{a,u}^H \mathbf{x}_a(t)} + n_{\mathrm{C} , u}(t),
	\label{eq:2-2}
\end{equation}
where $n_{\mathrm{C} , u}(t)$ indicates the complex additive white Gaussian noise (AWGN) at UE $u$.
The UE $u$ down-converts the received signal \eqref{eq:2-2} to baseband via the RF chains, and the baseband signal at $l$-th time slot is given by
\begin{equation}
	\begin{aligned}
		y_u[l] = & \sum_{a\in \mathcal{A}}{ \mathbf{h}_{a,u}^H \mathbf{F}_{\mathrm{A},a} \mathbf{f}_{\mathrm{D},a,u} s_{a,u}[l] } \\
		& + \sum_{a\in \mathcal{A}}{ \sum_{v \in \mathcal{U} , v \ne u} { \mathbf{h}_{a,u}^H \mathbf{F}_{\mathrm{A},a} \mathbf{f}_{\mathrm{D},a,v} s_{a,v}[l] } } + n_{\mathrm{C} , u} ,
	\end{aligned}
	\label{eq:2-3}
\end{equation}
where $n_{\mathrm{C} , u}\sim{\mathcal{CN}(0 , \sigma_{\mathrm{C},u}^2)}$ represents the complex AWGN with variance $\sigma_{\mathrm{C},u}^2$ at UE $u$.
According to \eqref{eq:2-3}, the achievable transmission rate at UE $u$ can be written as \cite{Barneto2022WCM}
\begin{equation}
	\begin{aligned}
		& \mathrm{Rate}_u \left( \{ \mathbf{F}_{\mathrm{A},a} \} , \{ \mathbf{F}_{\mathrm{D},a} \}\right) \\
		& = \log\left( 1 + \frac{\Big| \sum\limits_{a \in \mathcal{A}}{\mathbf{h}_{a,u}^H \mathbf{F}_{\mathrm{A},a} \mathbf{f}_{\mathrm{D},a,u}}  \Big|^2} { \sum\limits_{v \in \mathcal{U} , v \ne u}{\Big| \sum\limits_{a \in \mathcal{A}}{ \mathbf{h}_{a,u}^H \mathbf{F}_{\mathrm{A},a} \mathbf{f}_{\mathrm{D},a,v} } \Big|^2} + \sigma_{\mathrm{C},u}^2 } \right) .
	\end{aligned}
\end{equation}

\subsubsection{Radar Reception Model}
For clarity, we initially consider a simplified scenario where all APs cooperatively detect a single target in the presence of clutters. This framework is subsequently extended to the more general multi-target scenario.
To achieve this, each AP is equipped with a radar sensing receiver, which is co-located with the AP transmitter.
Based on above assumptions, the backscattered signal at sensing receiver in AP $a$ is given by
\begin{equation}
	\begin{aligned}
		& \mathbf{y}_{\mathrm{R},a}(t) = \sum\limits_{b \in \mathcal{A}}{ \xi_{a,b,0} \Upsilon_{a,b,0} \mathbf{G}_{a,b,0}\mathbf{F}_{\mathrm{A},a}\mathbf{F}_{\mathrm{D},a}\mathbf{s}_{b}(t-\tau_{a,b,0}) } \\
		& + \sum\limits_{q\in\mathcal{Q}}{ \sum\limits_{b \in \mathcal{A}}{ \xi_{a,b,q} \Upsilon_{a,b,q} \mathbf{G}_{a,b,q}\mathbf{F}_{\mathrm{A},a}\mathbf{F}_{\mathrm{D},a}\mathbf{s}_{b}(t-\tau_{a,b,q}) } } + \mathbf{n}_{\mathrm{R},a}(t),
	\end{aligned}
\end{equation}
where $\mathbf{n}_{\mathrm{R},a}(t)$ is the AWGN with power spectral density $\sigma_{\mathrm{S}}^2$.
$\mathbf{G}_{a,b,q}$$ = $$\mathbf{a}_\mathrm{R}(\varphi _{a,q}) \mathbf{a}_\mathrm{T}^H(\varphi_{b,q})$ is the effective radar channel from AP $b$$\to$Target($q$$=$$0$)/Clutter($q$$\in$$\mathcal{Q}$)$\to$AP $a$, with $\varphi_{a,q}$ being the direction-of-arrivals (DoAs) from AP $a$$\to$Target($q$$=$$0$)/Clutter($q$$\in$$\mathcal{Q}$).
$\tau_{a,b,q} = (\|\mathbf{i}_{\mathrm{A},a} - \mathbf{i}_{q}\|_F + \|\mathbf{i}_{\mathrm{A},b} - \mathbf{i}_{q}\|_F)/c$ is the time delay from AP $b$$\to$Target($q$$=$$0$)/Clutter($q$$\in$$\mathcal{Q}$)$\to$AP $a$, where $\mathbf{i}_{\mathrm{A},a}$ and $\mathbf{i}_{q}$ being the coordinate point of AP $a$ and target($q$$=$$0$)/clutter($q$$\in$$\mathcal{Q}$).
$\xi_{a,b,q}$ denotes the complex amplitude of the target($q$$=$$0$)/clutter($q$$\in$$\mathcal{Q}$) observed via the path AP $b$$\to$Target($q$$=$$0$)/Clutter($q$$\in$$\mathcal{Q}$)$\to$AP $a$.
$\Upsilon_{a,b,q}$ denotes the path loss of the target($q$$=$$0$)/clutter($q$$\in$$\mathcal{Q}$) observed via the path AP $b$$\to$Target($q$$=$$0$)/Clutter($q$$\in$$\mathcal{Q}$)$\to$AP $a$.
It is assumed that $\xi_{a,b,0}$ is considered deterministic but remains unknown, and $\xi_{a,b,q},q\in\mathcal{Q}$ follows a complex normal distribution $\xi_{a,b,q}\sim\mathcal{CN}(0,\varsigma_{a,b,q}^2),q\in\mathcal{Q}$.

Based on the above models, the workflow of cooperative sensing can be summarized in the following steps.

\textit{Step 1: Matched Filtering (MF).}
In this paper, we assume that all the AP sensing receivers are asynchronous in time, meaning that AP $a$ only has local knowledge about $\mathbf{F}_{\mathrm{A},a}$, $\mathbf{F}_{\mathrm{D},a}$ and $\mathbf{s}_a(t)$, and $\tau_{a,a,q}$ for $q\in{0,\mathcal{Q}}$.
Each AP $a$ performs MF processing based on $\tau_{a,a,0}$ and $\mathbf{s}_a(t)$. Therefore, the MF processing can be modeled as
\begin{equation}
	\begin{aligned}
		\mathbf{Y}_a = & \frac{1}{T_0}\int_{T_0}{ \mathbf{y}_{\mathrm{R},a}(t) \mathbf{s}_a^H(t - \tau_{a,a,0}) \text{d}t } \\
		=&  \xi_{a,a,0} \Upsilon_{a,a,0} \mathbf{G}_{a,a,0}\mathbf{F}_{\mathrm{A},a}\mathbf{F}_{\mathrm{D},a} \\
		& + \sum\limits_{q\in\mathcal{Q}}{ \xi_{a,a,q} \Upsilon_{a,a,q} \iota_{a,q} \mathbf{G}_{a,a,q}\mathbf{F}_{\mathrm{A},a}\mathbf{F}_{\mathrm{D},a} } + \tilde{\mathbf{N}}_{\mathrm{R},a} ,
	\end{aligned}
\end{equation}
where $\iota_{a,q}=R(\tau_{a,a,0}-\tau_{a,a,q})$ is the auto-correlation function with $R(0)=1$.
$\tilde{\mathbf{N}}_{\mathrm{R},a} = \frac{1}{T_0}\int_{T_0}{ \mathbf{n}_{\mathrm{R},a}(t) \mathbf{s}_a^H(t - \tau_{a,a,0}) \text{d}t }$ is the output AWGN with $\tilde{\mathbf{N}}_{\mathrm{R},a} = [\tilde{\mathbf{n}}_{\mathrm{R},a,1} , \cdots , \tilde{\mathbf{n}}_{\mathrm{R},a,U}]$ and $\tilde{\mathbf{n}}_{\mathrm{R},a,u} \sim \mathcal{CN}(\mathbf{0},\bar{\sigma}_\mathrm{R}^2)$. 
The equivalent noise power $\bar{\sigma}_\mathrm{R}^2$ can be expressed as $\bar{\sigma}_\mathrm{R}^2 = \frac{\sigma_{\mathrm{R}}^2}{{BT_0}}$, where $\sigma_{\mathrm{R}}^2=\sigma_{\mathrm{S}}^2 B$ and $BT_0$ are the radar noise power and time-bandwidth product, respectively.

\textit{Step 2: Receive Beamforming (RBF).}
Then, after vectoring $\mathbf{Y}_a$ and processing it by the RBF $\mathbf{w}_a$, the output is given by
\begin{equation}
	\begin{aligned}
		\bar{y}_a & = \mathbf{w}_a^H \mathsf{Vec}(\mathbf{Y}_a) = \mathbf{w}_a^H \hat{\mathbf{y}}_a \\
		&  = \xi_{a,a,0} \bar{x}_{a,a,0} + \sum\limits_{q\in\mathcal{Q}}{ \xi_{a,a,q} \bar{x}_{a,a,q} } + \bar{n}_{\mathrm{R},a} ,
	\end{aligned}
	\label{eq:2-7n}
\end{equation}
where $\bar{n}_{\mathrm{R},a} = \mathbf{w}_0^H \mathsf{Vec}(\tilde{\mathbf{N}}_{\mathrm{R},a})$.
$\bar{x}_{a,a,0} = \mathbf{w}_a^H \hat{\mathbf{G}}_{a,a,0} \mathbf{f}_a$ and $\bar{x}_{a,a,q} = \mathbf{w}_a^H \hat{\mathbf{G}}_{a,a,q} \mathbf{f}_a$ where $\hat{\mathbf{G}}_{a,a,0} = \mathbf{I}_U\otimes \Upsilon_{a,b,0}\mathbf{G}_{a,a,0}$ and $\hat{\mathbf{G}}_{a,a,q} = \mathbf{I}_U\otimes \Upsilon_{a,b,q} \iota_{a,q}\mathbf{G}_{a,a,q}$.

\textit{Step 3: Cooperative Sensing Detector.}
Each AP forwards the local information $\bar{y}_a$ to the CPU, where the CPU carries out the data fusion and cooperative sensing.
The cooperative sensing detector can be defined by the following proposition.
\begin{prop}\label{pro:new-1}
	In generalized likelihood ratio test (GLRT), the cooperative sensing detector can be given by
	\begin{equation}
		\varpi = \sum\limits_{a\in\mathcal{A}}{ \frac{|\bar{y}_a|^2}{ {\sigma_{\mathrm{E},a}^2} } }   \mathop \gtrless \limits_{\mathcal{H}_0}^{\mathcal{H}_1} \mathcal{T}
		\label{eq:2-8n}
	\end{equation}
	where $\sigma_{\mathrm{E},a}^2 = \bar{\mathbf{x}}_{a}^H\bm{\varsigma}_a\bar{\mathbf{x}}_{a} + \bar{\sigma}_{\rm{R}}^2\|\mathbf{w}_a\|_F^2$, $\bar{\mathbf{x}}_{a} = [\bar{x}_{a,a,1} ,$ $ \cdots , \bar{x}_{a,a,Q}]^T$,
	$\bm{\varsigma}_a = \mathsf{Diag}[\varsigma_{a,a,1} , \cdots , \varsigma_{a,a,Q}]^T$, and $\mathcal{T}$ is the detection threshold.
\end{prop}
\begin{IEEEproof}
Please refer to supplementary material (SM) Appendix \ref{adx:new-1}.	
\end{IEEEproof}

\begin{corollary}
	According to the above detector \eqref{eq:2-8n}, with given probability of false-alarm $\mathrm{Pr_{FA}}$, the probability of detection $\mathrm{Pr_D}$ can be derived as
	\begin{equation}
		\mathrm{Pr_D} = \mathbb{Q}_M^A\left( \sqrt{2\sum\limits_{a\in\mathcal{A}}{\mathrm{SINR}_a}} , \sqrt{F_{\chi_{(2A)}^{2}}^{-1}(1-\mathrm{Pr_{FA}})}  \right)
		\label{eq:2-9n}
	\end{equation}
	where $\mathbb{Q}_M^A$ is generalized Marcum Q-function with order $A$, $F_{\chi_{(A)}^{2}}^{-1}$ represents the inverse cumulative distribution function (CDF) of the chi-square distribution with order $A$, and the $\mathrm{SINR}_a$ is given by
	\begin{equation}
		\mathrm{SINR}_a = \frac{|\xi_{a,a,0}\bar{x}_{a,a,0}|^2}{\sigma_{\mathrm{E},a}^2} = \frac{|\mathbf{w}_a^H\hat{\mathbf{G}}_{a,a,0}\mathbf{f}_a|^2}{\sum\limits_{q\in\mathcal{Q}}{|\mathbf{w}_a^H\hat{\mathbf{G}}_{a,a,q}\mathbf{f}_a|^2} + \sigma_{\rm{R}}^2\|\mathbf{w}_a\|_F^2}
		\label{eq:2-10n}
	\end{equation}
\end{corollary}
\begin{IEEEproof}
	Please refer to SM Appendix \ref{adx:new-2}.	
\end{IEEEproof}

\begin{remark}[Design Insights]
	From \eqref{eq:2-9n}, we observe that the radar detection performance is a function of the sum radar SINR $\sum_{a\in\mathcal{A}}{\mathrm{SINR}_a}$, from which we derive the following insights:
	First, with a given probability of false-alarm $\mathrm{Pr_{FA}}$, the radar detection performance improves as the sum radar SINR increases.
	Therefore, we can improve the detection performance of considered Co-ISACNet by improving sum radar SINR.
	Second, employing more APs can achieve a higher sum radar SINR, thereby enhancing cooperative detection performance.
	This reveals that deploying more APs can significantly improve performance.
\end{remark}

\begin{remark}[Asynchronous Sensing]
	The proposed cooperative sensing approach is fully asynchronous, eliminating the need for highly synchronized clocks. 
	Specifically, in \textit{Step 1}, the $a$-th AP requires only local information for MF.
	Moreover, in \textit{Step 3}, the forwarding of local information to the CPU also operates asynchronously.
	This is because \eqref{eq:2-8n} represents the sum of the absolute values of local samples, known as non-coherent processing, which is not affected by time delays.
\end{remark}

\begin{remark}[Minimal Information Exchange]
	The proposed cooperative sensing approach requires only minimal information exchange. 
	Specifically, in \textit{Step 1}, the $a$-th AP needs only local information for MF. 
	Furthermore, in \textit{Step 3}, to achieve cooperative sensing, it is only necessary to forward the local information $| \bar{y}_a |^2 \in \mathbb{R}^1$ to the CPU.
\end{remark}

\begin{remark}[Scalability]
	Although the proposed cooperative sensing method is based on a single target scenario, it can be easily extended to multiple targets scenario. 
	Specifically, when detecting the $o$-th target among $O$ targets, the other targets should be treated as ``clutter sources". 
	Then, detecting $o$-th target can proceed using the same proposed cooperative sensing method \cite{Yu2020Multiple,Cheng2018Multiple}.
\end{remark}

\begin{figure}
	\centering
	\includegraphics[width=1\linewidth]{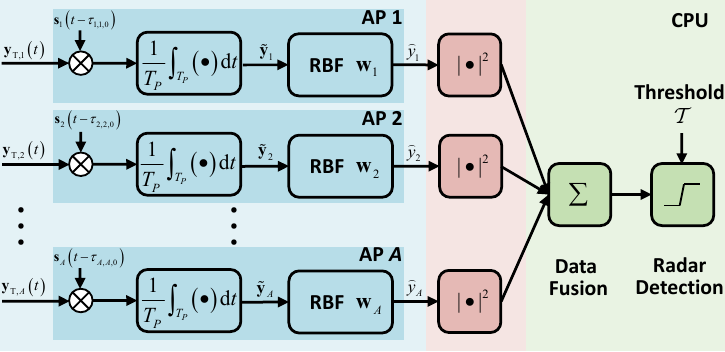}
	\vspace{-2em}
	\caption{The workflow of proposed cooperative sensing.}
	\vspace{-1em}
	\label{fig:Detector}
\end{figure}

\vspace{-1em}
\subsection{Problem Formulation}
\subsubsection{Performance Metrics}
For communication function, our objective is to enhance the overall communication capacity of the network. 
As demonstrated in subsection II-B.2, this can be achieved by maximizing the weighted sum rate (WSR), which is given by
\begin{equation}
	\mathrm{WSR}(\{\mathbf{F}_{\mathrm{A},a}\} , \{\mathbf{F}_{\mathrm{D},a}\}) =  \sum\limits_{u \in \mathcal{U}}{ {w_u}\mathrm{Rate}_u\left( \left\{ \mathbf{F}_{\mathrm{A},a} \right\} , \left\{ \mathbf{F}_{\mathrm{D},a} \right\} \right) }
\end{equation}
where $w_u \in \mathbb{R}^+$ is the weight of the UE $u$.

For radar function, as indicated in Remark 1, the sensing performance can be enhanced by improving the sum radar SINR.
However, we note that using radar SINR as a metric arouses the following drawbacks:
1) optimizing the sum radar SINR necessitates a joint transceiver design, which is often impractical in real-world scenarios.
2) optimizing sum radar SINR requires detailed knowledge of the target and clutter sources, which is typically challenging to obtain.

To address these drawbacks, we propose to design the transmit beampattern, a common method used in practical radar systems.
According to radar SINR \eqref{eq:2-10n}, the designed beampattern of the proposed Co-ISACNet system should have the following characteristics:
1) forming mainlobes towards targets;
2) achieving notch towards clutter sources;
3) achieving notch towards other APs\footnote{For a specific AP, the direct beams from other APs are relatively strong given the dense deployment of APs in mmWave frequencies, which causes the performance degradation of radar sensing. Therefore, the transmit beampattern is expected to limit the energy towards other APs.}.

To achieve the first characteristic of the desired transmit radar beampattern as a function of the detection angle $\theta$ at AP $a$, given by $\mathcal{P}_a \left( \mathbf{F}_{\mathrm{A},a} , \mathbf{F}_{\mathrm{D},a} , \theta \right) = \| \mathbf{a}_{\mathrm{T}}^H\left(\theta\right) \mathbf{F}_{\mathrm{A},a} \mathbf{F}_{\mathrm{D},a} \|_F^2$.
We propose to match it to a pre-defined spectrum $\mathbf{p}_a = [ P_a({\theta_1}), \cdots , P_a({\theta_L})]^T$, where $L$ denotes the number of the discrete grid points within the angle region $[-90^\circ , 90^\circ]$. 
The matching can be mathematically described by the weighted mean square error (MSE) between ${{\mathcal P}_a}\left( {{{\bf{F}}_{{\text{A}},a}},{{\bf{F}}_{{\text{D}},a}}},{\theta} \right)$ and $P_a\left( \theta \right)$ for beampattern approximation as
\begin{equation}
	\begin{aligned}
		\mathrm{MSE}_a & \left( \mathbf{F}_{\mathrm{A},a} , \mathbf{F}_{\mathrm{D},a} , {\Psi_a}  \right)  \\
		& = \frac{1}{L} \sum\limits_{l = 1}^L { \mu_l {{\left| {{{\mathcal P}_a}\left( \mathbf{F}_{\mathrm{A},a},\mathbf{F}_{\mathrm{D},a},{\theta_l} \right) - \Psi_a P_a\left( {{\theta _l}} \right)} \right|}^2}} ,
	\end{aligned}
	\label{eq:2-5}
\end{equation}
where $\Psi_a\in\mathbb{R}^+$ is a scaling parameter to be optimized\footnote{$\Psi_a$ is introduced to scale the normalized pre-defined spectrum $\mathbf{p}_a$, making the beampattern MSE more tractable \cite{liu2020joint}.}, $\mu_l$ is a predefined parameter to control the waveform similarity at $l$-th discrete spatial angle $\theta_l$.

To achieve the second and third characteristics of the desired transmit beampattern, we consider constraining the energy towards the notch region under a pre-defined threshold $\Gamma_a$, which yields 
$\max_{\vartheta_{a,t} \in {\bm \Theta}_a}  \mathcal{P}_a \left( \mathbf{F}_{\mathrm{A},a} , \mathbf{F}_{\mathrm{D},a} , \vartheta_{a,t} \right) \le \Gamma_a , \forall a$,
where ${\bm \Theta}_a  \in \mathbb{C}^{T_a \times 1} $ denotes the beampattern notch discrete grid angle set, with $T_a$ being the number of the discrete grid points within the notch region.

\subsubsection{Problem Statement}
Based on above illustrations, we aim to maximize the WSR of the proposed Co-ISACNet subject to the radar beampattern weighted MSE, the transmit power budget and the analog beamformer constraints.
Therefore, the joint transmit HBF design problem can be formulated as
\begin{subequations}
	\begin{align}
		&  \mathop {\max }\limits_{ \left\{ \mathbf{F}_{\mathrm{A},a} \right\} , \left\{ \mathbf{F}_{\mathrm{D},a} \right\} , {\bm \Psi} }  \mathrm{WSR}(\{\mathbf{F}_{\mathrm{A},a}\} , \{\mathbf{F}_{\mathrm{D},a}\}) \label{eq:2-7a} \\
		& \qquad \quad {\text{s.t.}} \quad  \mathrm{MSE}_a \left( \mathbf{F}_{\mathrm{A},a} , \mathbf{F}_{\mathrm{D},a} , {\Psi_a} \right) \le {\gamma _a},\forall a , \label{eq:2-7b} \\
		& \qquad \qquad\quad  \max_{\vartheta_{a,t} \in {\bm \Theta}_a} \mathcal{P}_a \left( \mathbf{F}_{\mathrm{A},a} , \mathbf{F}_{\mathrm{D},a} , \vartheta_{a,t} \right) \le \Gamma_a, \forall a , \label{eq:2-7c} \\
		& \qquad \qquad\quad \left\| \mathbf{F}_{\mathrm{A},a} \mathbf{F}_{\mathrm{D},a} \right\|_F^2 \le E,\forall a ,  \label{eq:2-7d}\\
		& \qquad \qquad\quad \left| {{{\left[ \mathbf{F}_{\mathrm{A},a} \right]}_{m,n}}} \right| = 1 ,\forall m,n,\forall a , \label{eq:2-7e}
	\end{align}
	\label{eq:2-7}%
\end{subequations}

The optimization problem \eqref{eq:2-7} is non-convex due to the log-fractional expression in the objective function, fourth-order constraint \eqref{eq:2-7b}, the constant modulus constraints of the analog beamformer and the coupling among variables.
In addition, the centralized optimization framework is unrealistic due to the unaffordable computational burden at the CPU, which needs to deal with extremely large dimension of HBFs. 
To tackle these two difficulties, in the following sections, we propose a distributed optimization framework which is suitable for solving general large-dimensional HBF design problems.

\section{Distributed Optimization Framework}\label{Sec-3}
In this section, we review the conventional centralized ADMM framework and propose a novel distributed optimization framework to solve the general HBF design problem for multi-AP scenarios.

We consider a multi-AP network design problem where APs equipped with HBF architectures collaborate to accomplish a certain task.
We write the multi-AP network optimization problems in the most general form as follows:
\begin{subequations}
	\begin{align}
		\min\limits_{ \{ \mathbf{F}_{\mathrm{A},a} \} , \{ \mathbf{F}_{\mathrm{D},a} \}} & \mathcal{G} \left(  \{ \mathbf{F}_{\mathrm{A},a} \} , \{ \mathbf{F}_{\mathrm{D},a} \}  \right) \label{eq:3-7a}\\
		\mathrm{s.t.} \qquad & {f}_i\left( \mathbf{F}_{\mathrm{A},a} , \mathbf{F}_{\mathrm{D},a} \right) = 0 , i \in \mathcal{I}_a, \forall a , \label{eq:3-7b}\\
		& {h}_j\left( \mathbf{F}_{\mathrm{A},a} , \mathbf{F}_{\mathrm{D},a} \right) \le 0 , j \in \mathcal{J}_a, \forall a , \label{eq:3-7c}\\
		& \mathbf{F}_{\mathrm{A},a} \in \mathcal{F} , \forall a \label{eq:3-7d},
	\end{align}
	\label{eq:3-7}%
\end{subequations}
where $\mathcal{I}_a=\{ 1 , \cdots , I_a \}$ and $\mathcal{J}_a=\{ 1 , \cdots , J_a \}$ collect the index of equality and inequality constraints, respectively. Here we state the properties that problem \eqref{eq:3-7} has as follows.

\textit{Property 1: Coupled Optimization Variables (Decision Variables).} $\mathbf{F}_{\mathrm{A},a}$ and $\mathbf{F}_{\mathrm{D},a}$ are analog and digital beamfomers at $a$-th AP, which are always coupled as $\mathbf{F}_{\mathrm{A},a}\mathbf{F}_{\mathrm{D},a}$ in objective \eqref{eq:3-7a} and constraints \eqref{eq:3-7b}-\eqref{eq:3-7c}.
	
\textit{Property 2: Structured Objective.} The objective function $ \mathcal{G} \left(  \{ \mathbf{F}_{\mathrm{A},a} \} , \{ \mathbf{F}_{\mathrm{D},a} \}  \right) $ is structured as a summation of a highly coupled component $\mathcal{G}_0 (  \{ \mathbf{F}_{\mathrm{A},a} \} , \{ \mathbf{F}_{\mathrm{D},a} \}  )$ and $A$ separated components $\mathcal{G}_a ( \mathbf{F}_{\mathrm{A},a} ,\mathbf{F}_{\mathrm{D},a} ), \forall a$, i.e.,
	\begin{equation}
		\begin{aligned}
			\mathcal{G} &  \left( \{ \mathbf{F}_{\mathrm{A},a} \} , \{ \mathbf{F}_{\mathrm{D},a} \} \right)  \\
			& = \mathcal{G}_0 \left( \{ \mathbf{F}_{\mathrm{A},a} \} , \{ \mathbf{F}_{\mathrm{D},a} \} \right) + \sum_{a \in \mathcal{A}} { \mathcal{G}_a \left( \mathbf{F}_{\mathrm{A},a} , \mathbf{F}_{\mathrm{D},a}\right) } ,
		\end{aligned}
	\end{equation}
where the coupled component $\mathcal{G}_0 (\mathbf{F})$ is convex, continuous and Lipchitz gradient continuous.

\textit{Property 3: Multiple Constraints.} Problem \eqref{eq:3-7} has equality constraints ${f}_i\left( \mathbf{F}_{\mathrm{A},a} , \mathbf{F}_{\mathrm{D},a} \right) = 0 , i \in \mathcal{I}_a, \forall a$,  inequality constraints ${h}_j\left( \mathbf{F}_{\mathrm{A},a} , \mathbf{F}_{\mathrm{D},a} \right) \le 0 , j \in \mathcal{J}_a, \forall a$ 
, and analog beamformer constraint $\mathbf{F}_{\mathrm{A},a} \in \mathcal{F} , \forall a$, where the set $\mathcal{F}$ is determined by the topology of the phase shift network.

The challenges for solving \eqref{eq:3-7} lie in the following three perspectives:
1) The HBF for the $a$-th AP, i.e., $\mathbf{F}_{\mathrm{A},a}$ and $\mathbf{F}_{\mathrm{D},a}$, are highly coupled in objective function $\mathcal{G} \left(  \{ \mathbf{F}_{\mathrm{A},a} \} , \{ \mathbf{F}_{\mathrm{D},a} \}  \right)$, equality constraints \eqref{eq:3-7b} and inequality constraints \eqref{eq:3-7c}.
2) The HBF for different APs are coupled in the objective function. In particular, if the objective function is complicated and non-convex, the problem \eqref{eq:3-7} would be extremely hard to solve.
3) Since the dimensions of analog beamformer $\mathbf{F}_{\mathrm{A},a} , \forall a$ and the number of APs $A$ are usually large, problem \eqref{eq:3-7} is a high-dimensional optimization problem, which results in high computational complexity.

\begin{remark}
	Our considered optimization problem \eqref{eq:3-7} differs from the existing decentralized consensus optimization problem (DCOP) \cite{chang2016asynchronous,chang2014multi,shi2015extra,shi2015proximal} in the following two aspects.
	1) In DCOP, each agent has its own private task, and each agent's decision affects other agents' tasks through variable coupling.
	However, the problem \eqref{eq:3-7} is a single-task optimization problem, where multiple agents (APs) collaborate to complete the same task.
	2) In DCOP, all the agents share the same decision variables.
	However, in problem \eqref{eq:3-7}, each AP has its own decision variables ($\mathbf{F}_{\mathrm{A},a}$ and $\mathbf{F}_{\mathrm{D},a}$), and the decision variables of different APs are coupled in the objective function.
\end{remark}

\vspace{-0.8em}
\subsection{Centralized ADMM Framework}

In this subsection, we review the conventional centralized optimization method to solve problem \eqref{eq:3-7}.
Centralized optimization usually transforms the original problem into multiple more tractable sub-problems.
Then, all the sub-problems are solved together in a CPU, so that the overall objective function is optimized.
Here, we introduce the centralized ADMM framework to solve problem \eqref{eq:3-7}.

To tackle the first challenge of solving problem \eqref{eq:3-7}, we  introduce auxiliary variables $\mathbf{T} = [ \mathbf{T}_1^H , \cdots ,\mathbf{T}_A^H ]^H$ satisfying $\mathbf{T}_a = \mathbf{F}_{\mathrm{A},a} \mathbf{F}_{\mathrm{D},a} , \forall a$ and convert the optimization problem as
\begin{subequations}
	\begin{align}
		\min\limits_{ \{ \mathbf{F}_{\mathrm{A},a} \} , \{ \mathbf{F}_{\mathrm{D},a} \} ,  \mathbf{T} } & \mathcal{G} \left( \mathbf{T}  \right) \\
		\mathrm{s.t.} \qquad\; & {f}_i\left( \mathbf{T}_a \right) = 0 , i \in  \mathcal{I}_a, \forall a , \label{eq:3-9b}\\
		& {h}_j\left( \mathbf{T}_a \right) \le 0 , j \in \mathcal{J}_a, \forall a , \label{eq:3-9c}\\
		& \mathbf{F}_{\mathrm{A},a} \in \mathcal{F} , \forall a  , \label{eq:3-9d} \\
		& \mathbf{T}_a = \mathbf{F}_{\mathrm{A},a} \mathbf{F}_{\mathrm{D},a}, \forall a . \label{eq:3-9e}
	\end{align}
	\label{eq:3-9}%
\end{subequations}
Following the ADMM framework, we penalize the equality constraints \eqref{eq:3-9d} into the objective function and  obtain the following augmented Lagrangian (AL) minimization problem
\begin{equation}
		\begin{aligned}
			\min\limits_{ \{ \mathbf{F}_{\mathrm{A},a} \} , \{ \mathbf{F}_{\mathrm{D},a} \} , \mathbf{T} } \; & \mathcal{L} \left( \{  \mathbf{F}_{\mathrm{A},a} \} , \{ \mathbf{F}_{\mathrm{D},a} \}, \mathbf{T} , \{ \mathbf{D}_a \} \right) \\
			\mathrm{s.t.} \qquad \;\; & \eqref{eq:3-9b} - \eqref{eq:3-9e} .
		\end{aligned}
	\label{eq:3-10}%
\end{equation}
where the scaled AL function is given by
\begin{equation}
	\begin{aligned}
		& \mathcal{L} \left( \{  \mathbf{F}_{\mathrm{A},a} \} , \{ \mathbf{F}_{\mathrm{D},a} \}, \mathbf{T}  \right) \\
		& \qquad = \mathcal{G} \left( \mathbf{T}  \right) + \sum_{a \in \mathcal{A}}{  \mathbb{I}_a \left(   \mathbf{F}_{\mathrm{A},a}  , \mathbf{F}_{\mathrm{D},a} , \mathbf{T}_a , \mathbf{D}_a \right) },
	\end{aligned}
	\label{eq:3-11}
\end{equation}
with $\mathbb{I}_a \left(   \mathbf{F}_{\mathrm{A},a}  , \mathbf{F}_{\mathrm{D},a} , \mathbf{T}_a , \mathbf{D}_a \right) =  \frac{\rho}{2}  \left\| \mathbf{T}_a - \mathbf{F}_{\mathrm{A},a}\mathbf{F}_{\mathrm{D},a} + \mathbf{D}_a \right\|_F^2 $, $\mathbf{D}_a$ the dual variable, and $\rho$ the penalty parameter.
Problem \eqref{eq:3-11} can be iteratively solved with the following centralized ADMM framework:
\begin{subequations}
	\begin{numcases}{}
		{\text{S1: }} \mathbf{T}^k = \arg \min_{\mathbf{T} \in \mathcal{X}} \mathcal{L} \left( \{  \mathbf{F}_{\mathrm{A},a}^{k-1} \} , \{ \mathbf{F}_{\mathrm{D},a}^{k-1} \}, \mathbf{T} , \{ \mathbf{D}_a^{k-1} \} \right) \notag \\
		{\text{S2: }} \{ \mathbf{F}_{\mathrm{A},a}^k \} = \mathop{\arg \min}\limits_{ \{ \mathbf{F}_{\mathrm{A},a} \} \in \mathcal{F} } \mathcal{L} \left( \{  \mathbf{F}_{\mathrm{A},a} \} , \{ \mathbf{F}_{\mathrm{D},a}^{k-1} \}, \mathbf{T}^k , \{ \mathbf{D}_a^{k-1} \}  \right)  \notag \\
		{\text{S3: }} \{ \mathbf{F}_{\mathrm{D},a}^k \} = \mathop{\arg \min}\limits_{ \{ \mathbf{F}_{\mathrm{D},a} \} } \mathcal{L} \left( \{  \mathbf{F}_{\mathrm{A},a}^k \} , \{ \mathbf{F}_{\mathrm{D},a} \}, \mathbf{T}^k  , \{ \mathbf{D}_a^{k-1} \}\right)  \notag \\
		{\text{S4: }} \text{Update Dual Variables: } \{\mathbf{D}_a^k\}  \notag
	\end{numcases}
\end{subequations}
where $\mathcal{X} = \mathcal{X}_1 \cup \mathcal{X}_2 \cdots \cup \mathcal{X}_A$ with $\mathcal{X}_a = \{ \mathbf{T}_a | {f}_i( \mathbf{T}_a ) = 0 , {h}_j( \mathbf{T}_a ) \le 0 , i \in \mathcal{I}_a , j \in \mathcal{J}_a\}$.
The dual variables are typically updated as $\mathbf{D}_a^k = \mathbf{D}_a^{k-1} + \mathbf{T}_a^k - \mathbf{F}_{\mathrm{A},a}^k \mathbf{F}_{\mathrm{D},a}^k , \forall a. $

Although the centralized ADMM can obtain satisfactory performance in many HBF design scenarios, it has many drawbacks when it comes to multi-AP network scenarios:
1) With high-dimensional optimization problems (S1-S3) for multi-AP networks with HBF architecture, the centralized ADMM framework is computationally expensive such that it cannot be practically utilized to solve problem \eqref{eq:3-11}.
2) Under the centralized ADMM framework, the CPU must have the information of all APs, such as CSI and parameter settings, which brings the CPU heavy tasks.
3) The centralized ADMM framework often requires much time to converge, which is unsuitable for real-time applications.
To address these issues, we will propose a distributed optimization framework with affordable computational complexity, reduced information exchange among APs, and fast convergence speed.

\vspace{-1em}
\subsection{Proposed Distributed Optimization Framework}
In this subsection, we go beyond the above centralized ADMM framework and propose a novel distributed optimization framework.
The core idea of distributed optimization is to decompose centralized objective function and constraints into distributed tasks such that they can be simultaneously implemented and the computational efficiency can be much increased.
However, problem \eqref{eq:3-7} cannot be distributively solved since each of S2-S4 can be further decoupled into $A$ distributed sub-problems, while S1 is coupled by the auxiliary variable $\mathbf{T}$.
To distributively solve problem \eqref{eq:3-7}, we propose a novel distributed optimization framework, namely \textbf{P}roximal gr\textbf{A}die\textbf{N}t \textbf{D}ecentralized \textbf{A}DMM (PANDA) framework, which modifies the centralized ADMM framework S1-S4 by decoupling S1 into $A$ sub-problems, each of which is the function of $\mathbf{T}_a$.
To do so, we first give the following lemma to find a surrogate problem of S1 based on PG. 
\begin{lemma}\label{lem:1}
	Let $\mathcal{G}(\mathbf{T})$ be a continuously differentiable function.
	For all $\mathbf{T}$,  the following inequality \cite{bertsekas1997nonlinear} holds:
	\begin{equation}
		\begin{aligned}
			\mathcal{G} \left(\mathbf{T}\right) 
			\le & \mathcal{G}\left(\mathbf{T}^k\right)  
			+ \frac{\alpha}{2} \left\|\mathbf{T}-\mathbf{T}^k\right\|_F^2 \\
			& + \Re \left\{ \mathsf{Tr} \left( \nabla \mathcal{G} \left(\mathbf{T}^k\right)^H \left(\mathbf{T}-\mathbf{T}^k\right) \right) \right\} ,
		\end{aligned}
	\end{equation}
	where $\mathbf{T}^k$ is the point at the $k$-th iteration, and $\alpha$ is the Lipschitz constant.
\end{lemma}
\begin{IEEEproof}
	Please refer to \cite[Descent Lemma]{bertsekas1997nonlinear}.
\end{IEEEproof}	

Performing \textit{Lemma \ref{lem:1}} to the coupled part $\mathcal{G}_0 (\mathbf{T})$ of the objective in S1, we replace $\mathcal{L} ( \{  \mathbf{F}_{\mathrm{A},a}^{k-1} \} , \{ \mathbf{F}_{\mathrm{D},a}^{k-1} \}, \mathbf{T} , \{ \mathbf{D}_a^{k-1} \} )$ with its locally tight upper bound function at $\mathbf{T}^k$ and obtain the following update
\begin{equation}
	\begin{aligned}
		\mathbf{T}^k = &  \arg \min_{\mathbf{T} \in \mathcal{X}} \Big\{ \Re \left\{ \mathsf{Tr} \left( \nabla \mathcal{G}_0 (\mathbf{T}^{k-1})^H (\mathbf{T} - \mathbf{T}^{k-1}) \right) \right\}  \\
		& \qquad \quad + \frac{\alpha}{2} \left\|  \mathbf{T} - \mathbf{T}^{k-1} \right\|_F^2  + \sum_{a \in \mathcal{A}} { \mathcal{G}_a \left( \mathbf{T}_a \right) } \\
		& \qquad \quad + \frac{\rho}{2} \sum_{a \in \mathcal{A}}{  \left\| \mathbf{T}_a - \mathbf{F}_{\mathrm{A},a}^{k-1}\mathbf{F}_{\mathrm{D},a}^{k-1} + \mathbf{D}_a^{k-1} \right\|_F^2 } \Big\}
	\end{aligned}
	\label{eq:3-14}
\end{equation}
where $\nabla \mathcal{G}_0 (\mathbf{T}) = [\nabla \mathcal{G}_0^H (\mathbf{T}_1) , \cdots , \nabla \mathcal{G}_0^H (\mathbf{T}_A)]^H$ is the first-order derivative, 
$\alpha$ is the Lipschitz constant of $\mathcal{G}_0 (\mathbf{T})$.
Now, the objective function and constraints are separable, which means updating \eqref{eq:3-14} can be further decoupled as
\begin{equation}
	\begin{aligned}
		\mathbf{T}_a^k = &  \arg \min_{\mathbf{T}_a \in \mathcal{X}_a} \Big\{ \Re \left\{ \mathsf{Tr} \left( \nabla \mathcal{G}_0 (\mathbf{T}_a^{k-1})^H (\mathbf{T}_a -\mathbf{T}_a^{k-1}) \right) \right\}  \\
		& \qquad \qquad + \frac{\alpha}{2} \left\|  \mathbf{T}_a - \mathbf{T}_a^{k-1} \right\|_F^2  + \mathcal{G}_a \left( \mathbf{T}_a \right) \\
		& \qquad \qquad + \frac{\rho}{2}   \left\| \mathbf{T}_a - \mathbf{F}_{\mathrm{A},a}^{k-1}\mathbf{F}_{\mathrm{D},a}^{k-1} + \mathbf{D}_a^{k-1} \right\|_F^2 \Big\} \\
		= & \mathrm{Prox}_{\mathcal{G}_a , \beta}^{\mathcal{X}_a} \left[ \frac{1}{\beta}  \nabla \widetilde{\mathcal{L}}_a \left( {\mathbf{T}}_a^{k - 1} , {\mathbf{F}}_{{\text{A}},a}^{k - 1} , {\mathbf{F}}_{{\text{D}},a}^{k - 1} \right) \right].
	\end{aligned}
	\label{eq:3-15}
\end{equation}
where 
$\nabla \widetilde{\mathcal{L}}_a ( {\mathbf{T}}_a^{k - 1} , {\mathbf{F}}_{{\text{A}},a}^{k - 1} , {\mathbf{F}}_{{\text{D}},a}^{k - 1} , {\mathbf{D}}_a^{k - 1} ) = -\nabla {\mathcal{G}_0}({\mathbf{T}}_a^{k - 1}) + {\alpha}{\mathbf{T}}_a^{k - 1} + \rho ( {\mathbf{F}}_{{\text{A}},a}^{k - 1}{\mathbf{F}}_{{\text{D}},a}^{k - 1} - {\mathbf{D}}_a^{k - 1} )$,
and $\beta = \alpha + \rho$.
The PG operator for function $\mathcal{G} (\mathbf{T})$ at a given point $\widehat{\mathbf{T}}$ is given by
$\mathrm{Prox}_{\mathcal{G} , \beta}^{\mathcal{X}} \big[ \widehat{\mathbf{T}} \big] = \arg \min_{\mathbf{T} \in \mathcal{X}} \mathcal{G}(\mathbf{T}) + \frac{\beta}{2} \big\| \mathbf{T} - \widehat{\mathbf{T}} \big\|_F^2$.

Finally, the proposed PANDA framework consists of the following iterative steps
\begin{subequations}
	\begin{numcases}{}
		{\text{P1: }} \mathbf{T}_a^k  = \mathrm{Prox}_{\mathcal{G}_a , \beta}^{\mathcal{X}_a} \left[ \frac{1}{\beta}  \nabla \widetilde{\mathcal{L}}_a \left( {\mathbf{T}}_a^{k - 1} , {\mathbf{F}}_{{\text{A}},a}^{k - 1} , {\mathbf{F}}_{{\text{D}},a}^{k - 1} , {\mathbf{D}}_a^{k - 1} \right) \right] \nonumber \\
		{\text{P2: }} \mathbf{F}_{\mathrm{A},a}^k  = \arg \min_{\mathbf{F}_{\mathrm{A},a}  \in \mathcal{F} } \mathbb{I}_a \left(   \mathbf{F}_{\mathrm{A},a}  , \mathbf{F}_{\mathrm{D},a}^{k-1} , \mathbf{T}_a^{k} , \mathbf{D}_a^{k-1} \right) \nonumber\\
		{\text{P3: }} \mathbf{F}_{\mathrm{D},a}^k  = \arg \min_{\mathbf{F}_{\mathrm{D},a}} \mathbb{I}_a \left(   \mathbf{F}_{\mathrm{A},a}^{k}  , \mathbf{F}_{\mathrm{D},a} , \mathbf{T}_a^{k} , \mathbf{D}_a^{k-1} \right) \nonumber\\
		{\text{P4: }} \text{Update Dual Variables: } \{\mathbf{D}_a^k\} \nonumber
	\end{numcases}
\end{subequations}
By applying the PANDA framework, P1-P4 can be distributively solved at each AP with local information, which significantly improves the beamforming design efficiency compared to the conventional centralized optimization methods.

\begin{lemma}\label{lem:2}
	The sequence $\left\{ \left\{ \mathbf{T}_a^k \right\} , \left\{ \mathbf{F}_{\mathrm{A},a}^k \right\} , \left\{ \mathbf{F}_{\mathrm{D},a}^k \right\}  \right\}$ generated by the proposed PANDA has the following properties:
	\begin{enumerate}[leftmargin=13pt]
		\item The proposed PANDA framework modifies S1 in the centralized ADMM without affecting the monotonicity.
		
		\item Under the mild conditions $\lim_{k\to \infty} \mathbf{D}_a^{k+1} - \mathbf{D}_a^k = \mathbf{0}, \forall a$, there exists an stationary point $\left\{ \left\{ \mathbf{T}_a^\star \right\} , \left\{ \mathbf{F}_{\mathrm{A},a}^\star \right\} , \left\{ \mathbf{F}_{\mathrm{D},a}^\star \right\}  \right\}$, which is the optimal solution to \eqref{eq:3-7}.
	\end{enumerate}
\end{lemma}
\begin{IEEEproof}
	Please refer to SM Appendix \ref{adx:1}.
\end{IEEEproof}

\begin{remark}
	The proposed framework is different from the existing PG based decentralized ADMM \cite{shi2015extra,shi2015proximal,chang2014multi} from the following two aspects.
	1) Existing papers \cite{shi2015extra,shi2015proximal,chang2014multi} adopt PG to approximate the non-smooth part in objective function to reduce the complexity.
	However, we adopt PG to decouple the objective function into independent parts to facilitate distributed optimization. 
	2) Instead of centrally optimizing overall decision variables ($\{\mathbf{F}_{\mathrm{A},a}\}$, $\{\mathbf{F}_{\mathrm{D},a}\}$) for all agents (APs) in a CPU, our proposed framework optimizes the sub-set of decision variables ($\mathbf{F}_{\mathrm{A},a}$, $\mathbf{F}_{\mathrm{D},a}$) in corresponding agent (AP), which decreases the computational complexity and reduces backhaul signaling.
\end{remark}

In the next section, we will customize the proposed PANDA framework to distributively solve problem \eqref{eq:2-7}.

\section{Distributed Optimization to Problem \eqref{eq:2-7}}\label{Sec-4}

In this section, we reformulate problem \eqref{eq:2-7} to facilitate the use of the proposed PANDA framework and illustrate the  solution of \eqref{eq:2-7} in detail.

\vspace{-1em}
\subsection{The PANDA Framework of Problem \eqref{eq:2-7}}

\subsubsection{Problem Reformulation}

The objective \eqref{eq:2-7a} is the sum of non-convex logarithmic functions, which complicates the design and hinders the employment of the proposed PANDA framework.
Additionally, the quartic radar MSE constraints in  \eqref{eq:2-7b} also present challenges in employing the proposed PANDA framework. 
Therefore, before developing the PANDA framework for problem \eqref{eq:2-7}, we propose the following problem transformation.

\textit{Setp 1, Objective Transformation:} We first deal with the objective function with the following proposition.
\begin{prop}\label{Pro:1}
	Exploiting fractional programming and introducing auxiliary variables ${\bf r} = [r_1 , \cdots , r_U]^T$ and ${\bm \eta} = [\eta_1 , \cdots , \eta_U]^T$, objective \eqref{eq:2-7a} can be reformulated as
	\begin{subequations}
		\begin{align}
			& \mathcal{G}  \left( \left\{ \mathbf{F}_{\mathrm{A},a} \right\} , \left\{ \mathbf{F}_{\mathrm{D},a} \right\} \right)\nonumber \\ &= -\sum\limits_{u \in \mathcal{U}}{ {w_u}\mathrm{Rate}_u\left( \left\{ \mathbf{F}_{\mathrm{A},a} \right\} , \left\{ \mathbf{F}_{\mathrm{D},a} \right\} \right) }\\
			& = - \sum\limits_{u \in \mathcal{U}} {2\sqrt {{w_u}\left( {1 + {r_u}} \right)} } \Re \{ {\eta _u^*\sum\limits_{a \in \mathcal{A}} { [ \mathbf{H}_a^H \mathbf{F}_{\mathrm{A},a} \mathbf{F}_{\mathrm{D},a} ]_{u,u} } } \} \nonumber \\ 
			& \quad + \sum\limits_{u \in \mathcal{U}} {{{\left| {{\eta _u}} \right|}^2} \Big( {\sum\limits_{v \in \mathcal{U}} {{{\big| {\sum\limits_{a \in \mathcal{A}} { [ \mathbf{H}_a^H \mathbf{F}_{\mathrm{A},a} \mathbf{F}_{\mathrm{D},a} ]_{u,v} } } \big| }^2}}  + \sigma_{\mathrm{C},u}^2} \Big)} \nonumber \\
            & \quad - \sum\limits_{u \in \mathcal{U}} {{w_u} \log \left( {1 + {r_u}} \right)}  + \sum\limits_{u \in \mathcal{U}} {{w_u}{r_u}}, \label{eq:3-30a} \\
			& = \underbrace{\left\| \mathbf{B}_1\mathbf{F} \right\|_F^2}_{\mathcal{G}_0 \left( \{ \mathbf{F}_{\mathrm{A},a} \} , \{ \mathbf{F}_{\mathrm{D},a} \} \right) } + \sum_{a \in \mathcal
			A}{ \underbrace{\Re \left\{ {{\mathsf{Tr}} \left[ \mathbf{B}_{2,a} \mathbf{F}_{\mathrm{A},a} \mathbf{F}_{\mathrm{D},a} \right]} \right\}}_{ \mathcal{G}_a \left( \mathbf{F}_{\mathrm{A},a} , \mathbf{F}_{\mathrm{D},a} \right) } }  \nonumber \\
			& ~~~\quad - \sum_{u \in \mathcal{U}}{\mathrm{c}_{1,u}(r_u,\eta_u)} , 
			\label{eq:4-16b}
		\end{align}
		\label{eq:4-16}%
	\end{subequations}
	where the fresh notations are defined as
	\begin{subequations}
		\begin{align}
			& \mathbf{F} = [\mathbf{F}_{\mathrm{D},1}^H \mathbf{F}_{\mathrm{A},1}^H , \; , 
			  \mathbf{F}_{\mathrm{D},A}^H \mathbf{F}_{\mathrm{A},A}^H ]^H, \quad \mathbf{J}_1 = \mathsf{Diag}( \eta_1^* , \cdots , \eta_U^*) , \notag\\
			& \mathbf{J}_2 = \mathsf{Diag}( \sqrt {{w_1}( {1 + {r_1}} )} {\eta_1^*}, \cdots ,\sqrt {{w_U}( {1 + {r_U}} )} {\eta_U^*} ) \notag\\
			& \mathbf{B}_1 = \mathbf{J}_1 [ \mathbf{H}_1^H , \cdots , \mathbf{H}_A^H ] , \quad 
			  \mathbf{B}_2 = -2\mathbf{J}_a \mathbf{H}_a^H \notag\\
			& \mathrm{c}_{1,u}(r_u,\eta_u) = {{w_u} \log ( {1 + {r_u}} )}  - {{w_u}{r_u}} - {{\left| {{\eta _u}} \right|}^2 \sigma_{\mathrm{C},u}^2} .  \notag
		\end{align}
	\end{subequations}
\end{prop}
\begin{IEEEproof}
	Please refer to \cite[Appendix A]{li2019hybrid}.
\end{IEEEproof}

After applying \textit{Proposition \ref{Pro:1}}, $\mathcal{G} ( \left\{ \mathbf{F}_{\mathrm{A},a} \right\},\left\{ \mathbf{F}_{\mathrm{D},a} \right\} )$ is convex with respect to each variable with fixed others.

\textit{Step 2: Radar MSE Simplification:} To deal with the quartic constraint \eqref{eq:2-7a}, we present the following proposition.
\begin{prop}\label{Pro:2}
	The quartic optimization $\min_{{\bf X}}  \| {\bf XX}^H - N{\bf I} \|_F^2$ can be inexactly solved by optimizing the following simpler quadratic problem.
	\begin{equation}
		\min\limits_{{\bf X},{\bf Z}} \left\| {\bf X} - \sqrt{N}{\bf Z} \right\|_F^2 \qquad {\rm s.t.}\;  {\bf Z}^H{\bf Z} = {\bf I}, \label{eq:4-25N}
	\end{equation}
	where ${\bf Z}$ is a semiunitary matrix.
\end{prop}
\begin{IEEEproof}
	Please refer to \cite[Sec. IV]{tropp2005designing}.
\end{IEEEproof}	

By plugging $\mathbf{x}= \mathbf{F}_{\mathrm{D},a}^H \mathbf{F}_{\mathrm{A},a}^H {\bf{a}}_T\left( {{\theta _l}} \right) $, $\mathbf{z}^H\mathbf{z}=P(\theta_l)$ and $N=\Psi_a$ to \eqref{eq:4-25N} in \textit{Proposition \ref{Pro:2}}, we reformulate the radar beampattern weighted MSE as
\begin{equation}
	\begin{aligned}
		& \overline{\mathrm{MSE}}_a \left( \mathbf{F}_{\mathrm{A},a} , \mathbf{F}_{\mathrm{D},a} ,\mathbf{V}_a , {\zeta_a} \right) \\
		& \quad = \frac{1}{L}\sum\limits_{l = 1}^L {\mu_{a,l}\left\| {{\bf{a}}_T^H\left( {{\theta _l}} \right)\mathbf{F}_{\mathrm{A},a} \mathbf{F}_{\mathrm{D},a} - {\zeta_a}{\bf{v}}_{a,l}^H} \right\|_F^2} \le \gamma_a,
	\end{aligned}
\end{equation}
where $\mathbf{V}_a=[\mathbf{v}_{a,1},\cdots,\mathbf{v}_{a,L}] \in \mathbb{C}^{U \times L} , \forall a$ is the auxiliary variables, satisfying $\| {{{\bf{v}}_{a,l}}} \|_F^2 = P_a({\theta_l}) , \forall a,l$, and ${\zeta_a} = \sqrt{\Psi_a}$.

\subsubsection{Application of PANDA Framework}
With the above reformulation, the joint design problem \eqref{eq:2-7} can be recast as 
\begin{subequations}
	\begin{align}
		&\mathop {\min }\limits_{ \left\{ \mathbf{F}_{\mathrm{A},a} \right\} , \left\{ \mathbf{F}_{\mathrm{D},a} \right\} , {\bm \zeta}, \mathbf{r} , {\bm \eta} , \{\mathbf{V}_a\}}   \mathcal{G} \left( \left\{ \mathbf{F}_{\mathrm{A},a} \right\} , \left\{ \mathbf{F}_{\mathrm{D},a} \right\}  \right) \label{eq:4-19a} \\
		& \quad \qquad\;\;  {\text{s.t.}} \;\;\;   \overline{\mathrm{MSE}}_a \left( \mathbf{F}_{\mathrm{A},a} , \mathbf{F}_{\mathrm{D},a} , \mathbf{V}_a , {\zeta_a} \right) \le {\gamma _a},\forall a , \label{eq:4-19b} \\
		& \quad \qquad \qquad \; \max_{\vartheta_{a,t} \in {\bm \Theta}_a} \mathcal{P}_a \left( \mathbf{F}_{\mathrm{A},a} , \mathbf{F}_{\mathrm{D},a} , \vartheta_{a,t} \right) \le \Gamma_a, \forall a , \label{eq:4-19c} \\
		& \quad \qquad \qquad \; \left\| \mathbf{F}_{\mathrm{A},a} \mathbf{F}_{\mathrm{D},a} \right\|_F^2 = E,\forall a ,  \label{eq:4-19d}\\
		& \quad \qquad \qquad \; \left| {{{\left[ \mathbf{F}_{\mathrm{A},a} \right]}_{m,n}}} \right| = 1 ,\forall m,n,\forall a . \label{eq:4-19e}
	\end{align}
	\label{eq:4-19}%
\end{subequations}
Now, the challenge for solving \eqref{eq:4-19} lies in the coupling of $\mathbf{F}_{\mathrm{A},a}$ and $\mathbf{F}_{\mathrm{D},a}$ in the objective and constraints \eqref{eq:4-19b}-\eqref{eq:4-19e}.
To decouple $\mathbf{F}_{\mathrm{A},a}$ and $\mathbf{F}_{\mathrm{D},a}$ within and among \eqref{eq:4-19b}-\eqref{eq:4-19e}, we introduce several linear constraints $\mathbf{T}_a = \mathbf{U}_a = \mathbf{F}_{\mathrm{A},a} \mathbf{F}_{\mathrm{D},a} , \forall a$ and $\mathbf{z}_{a,t}^H = \mathbf{a}_\mathrm{T}\left( \vartheta_t\right) \mathbf{F}_{\mathrm{A},a} \mathbf{F}_{\mathrm{D},a}, \forall a,t$, penalize each of them, and formulate the AL minimization problem as
\begin{subequations}
	\begin{align}
		\mathop {\min }\limits_{\substack{ \left\{ \mathbf{F}_{\mathrm{A},a} \right\} , \left\{ \mathbf{F}_{\mathrm{D},a} \right\}, {\bm \Psi}, \mathbf{r} , {\bm \eta} , \\  \left\{ \mathbf{T}_a \right\} , \left\{ \mathbf{U}_a \right\} , \left\{ \mathbf{Z}_{a} \right\} , \{\mathbf{V}_a\}}  }  
		& {\Large \substack{\mathcal{L} (  \{ \mathbf{T}_a \} , \{ \mathbf{U}_a \} , \{ \mathbf{Z}_{a} \} , \qquad \quad \\
		\{ \mathbf{F}_{\mathrm{A},a} \} , \{ \mathbf{F}_{\mathrm{D},a} \} , \{ \mathbf{D}_a \} ) } } \label{eq:4-20a} \\
		{\text{s.t.}} \qquad \quad  & \overline{\mathrm{MSE}}_a \left( \mathbf{U}_a ,  \mathbf{V}_a , {\zeta_a} \right) \le {\gamma _a},\forall a , \label{eq:4-20b} \\
		& \max_{t_a} \left\| \mathbf{z}_{a,t_a} \right\|_F^2 \le \Gamma_a, \forall a ,\label{eq:4-20c} \\
		& \left\| \mathbf{T}_a \right\|_F^2 = E,\forall a ,  \label{eq:4-20d}\\
		& \left| {{{\left[ \mathbf{F}_{\mathrm{A},a} \right]}_{m,n}}} \right| = 1 ,\forall m,n,\forall a , \label{eq:4-20e}
	\end{align}
	\label{eq:4-20}%
\end{subequations}
where $t_a \in [1,\cdots,T_a]^T$, $\mathbf{Z}_a = [\mathbf{z}_{a,1} , \cdots , \mathbf{z}_{a,T_a}]$.
The AL function is defined as
\begin{equation}
	\begin{aligned}
		& \mathcal{L}  \left( \{ \mathbf{T}_a \} , \{ \mathbf{U}_a \} , \{ \mathbf{Z}_{a} \} , \{ \mathbf{F}_{\mathrm{A},a} \} , \{ \mathbf{F}_{\mathrm{D},a} \}  , \{ \mathbf{D}_a \} \right) \\
		& = \mathcal{G} \left( \{ \mathbf{T}_a \}  \right) + \sum_{a \in \mathcal{A}}{ \mathbb{I}_a ( \mathbf{T}_a , \mathbf{U}_a , \mathbf{Z}_{a} , \mathbf{F}_{\mathrm{A},a} , \mathbf{F}_{\mathrm{D},a} , \mathbf{D}_a ) } , 
  \end{aligned}
\end{equation}
where $\mathbf{D}_a$ is a collection of all dual variables, i.e., $\mathbf{D}_a = \{ {\bm \Omega}_a, {\bm \Lambda}_a, {\bm \Phi}_a \}$.
$\mathbb{I}_a ( \mathbf{T}_a , \mathbf{U}_a , \mathbf{Z}_{a} , \mathbf{F}_{\mathrm{A},a} , \mathbf{F}_{\mathrm{D},a} , \mathbf{D}_a ) = \frac{\rho}{2}\| \mathbf{T}_a - \mathbf{F}_{\mathrm{A},a} \mathbf{F}_{\mathrm{D},a} + {\bm \Omega}_a \|_F^2 + \frac{\varrho}{2}\| \mathbf{U}_a - \mathbf{F}_{\mathrm{A},a} \mathbf{F}_{\mathrm{D},a} + {\bm \Lambda}_a \|_F^2 + \frac{\lambda}{2}\| \mathbf{Z}_{a} - \mathbf{F}_{\mathrm{D},a}^H \mathbf{F}_{\mathrm{A},a}^H \mathbf{A}_{\mathrm{N},a} + {\bm \Phi}_{a} \|_F^2$ 
with $\mathbf{A}_{\mathrm{N},a} = [\mathbf{a}_\mathrm{T}(\vartheta_1) , \cdots , \mathbf{a}_\mathrm{T}(\vartheta_{T_a})]$.

Now we can adopt the proposed distributed framework following the steps in the sequel. 
\begin{subequations}
	\begin{numcases}{}
		{\overline{\text{P1}}:\;} \mathbf{T}_a^k  = \mathrm{Prox}_{\mathcal{G}_a , \beta}^{\mathcal{X}_{1,a}} \Big[ \frac{1}{\beta} \nabla \widetilde{\mathcal{L}}_a  \left(  \mathbf{T}_a^{k-1} ,  \mathbf{F}_{\mathrm{A},a}^{k-1} , \mathbf{F}_{\mathrm{D},a}^{k-1} , {{\bm \Omega}}_a^{k - 1} \right)  \Big] \nonumber\\
		{\overline{\text{P2}}:\;} \mathbf{U}_a^k   = \mathop{\mathrm{arg \; min}}\limits_{\mathbf{U}_{a} \in \mathcal{X}_{2,a} }  \| \mathbf{U}_a - \mathbf{F}_{\mathrm{A},a}^{k-1} \mathbf{F}_{\mathrm{D},a}^{k-1} + {\bm \Lambda}_a^{k-1} \|_F^2   \nonumber\\
		{\overline{\text{P3}}:\;} \mathbf{Z}_{a}^k =  \mathop{\mathrm{arg \; min}}\limits_{\mathbf{Z}_{a} \in \mathcal{X}_{3,a}}\| \mathbf{Z}_{a} - (\mathbf{F}_{\mathrm{A},a}^{k-1} \mathbf{F}_{\mathrm{D},a}^{k-1})^H \mathbf{A}_\mathrm{N} + {\bm \Phi}_{a}^{k-1} \|_F^2 \nonumber  \\
		{\overline{\text{P4}}:\;} \mathbf{F}_{\mathrm{A},a}^k  = \mathop{\mathrm{arg \; min}}\limits_{\mathbf{F}_{\mathrm{A},a} \in \mathcal{X}_{4,a}} \mathbb{I}_a  ( \mathbf{T}_a^k  ,  \mathbf{U}_a^k , \mathbf{Z}_{a}^k , \mathbf{F}_{\mathrm{A},a}  ,  \mathbf{F}_{\mathrm{D},a}^{k-1} ) \nonumber \\
		{\overline{\text{P5}}:\;} \mathbf{F}_{\mathrm{D},a}^k  = \mathop{\mathrm{arg \; min}}\limits_{\mathbf{F}_{\mathrm{D},a}}  \mathbb{I}_a  \left( \mathbf{T}_a^k  ,  \mathbf{U}_a^k , \mathbf{Z}_{a}^k , \mathbf{F}_{\mathrm{A},a}^k  ,  \mathbf{F}_{\mathrm{D},a} \right) \nonumber \\
		{\overline{\text{P6}}:\;} \text{Update Dual Variables: } \{ {\mathbf{D}}_a \} \nonumber
	\end{numcases}
\end{subequations}
where $\beta = \alpha + \rho $ and $\nabla \widetilde{\mathcal{L}}_a ( {\mathbf{T}}_a^{k - 1} , {\mathbf{F}}_{{\text{A}},a}^{k - 1} , {\mathbf{F}}_{{\text{D}},a}^{k - 1} , {\bm \Omega}_a^{k - 1} ) = -\nabla {\mathcal{G}_0}({\mathbf{T}}_a^{k - 1}) + {\alpha}{\mathbf{T}}_a^{k - 1} + \rho ( {\mathbf{F}}_{{\text{A}},a}^{k - 1}{\mathbf{F}}_{{\text{D}},a}^{k - 1} - {\bm \Omega}_a^{k - 1} )$ with $\nabla {\mathcal{G}_0}({\mathbf{T}}_a^{k - 1})  
=  \mathbf{H}_a \mathbf{J}_1^H \mathbf{J}_1 ( \sum_{i \in \mathcal{A}}{{\bm \Xi}_i^{k-1}} )$ and ${\bm \Xi}_a = \mathbf{H}_a^H \mathbf{T}_a$.
$\mathcal{X}_{1,a} = \{ \mathbf{T}_a | \| \mathbf{T}_a \|_F^2 = E \}$,
$\mathcal{X}_{2,a} = \{ \mathbf{U}_a  | \overline{\mathrm{MSE}}_a ( \mathbf{U}_a ,  \mathbf{V}_a , \zeta_a ) $ $\le {\gamma _a} \}$,
$\mathcal{X}_{3,a} = \{ \mathbf{Z}_{a}  | \max_{t_a } \| \mathbf{z}_{a,t_a} \|_F^2 \le \Gamma_a \}$, and
$\mathcal{X}_{4,a} = \{ \mathbf{F}_{\mathrm{A},a}  | | {{{\left[ \mathbf{F}_{\mathrm{A},a} \right]}_{m,n}}} | = 1 ,\forall m,n \}$.
In what follows, we discuss respectively the solutions to ${\overline{\text{P1}}}$-${\overline{\text{P5}}} $.

\vspace{-1em}
\subsection{Solution to Sub-problems}

\subsubsection{Update $\mathbf{T}_a$}
Given other variables, $\mathbf{T}_a$ can be updated by PG method as
\begin{equation}
	\mathbf{T}_a^k  = \mathrm{Prox}_{\mathcal{G}_a , \beta}^{\mathcal{X}_{1,a}} \Big[ \frac{1}{\beta} \nabla \widetilde{\mathcal{L}}  \left(  \mathbf{T}_a^{k-1} ,  \mathbf{F}_{\mathrm{A},a}^{k-1} , \mathbf{F}_{\mathrm{D},a}^{k-1} , {\bm \Omega}_a^{k - 1} \right)  \Big] ,
	\label{eq:4-24}
\end{equation}
According to the definition of PG operation, problem \eqref{eq:4-24} can be equivalently rewritten as
\begin{equation}
	\begin{aligned}
		\mathop {\min }\limits_{\mathbf{T}_a} \; & \Re\{ \mathsf{Tr} ( {\mathbf{B}}_2  \mathbf{T}_{a} )\} + \frac{\beta}{2} \Big\| \mathbf{T}_a  -  \frac{1}{\beta} \big( -\nabla {\mathcal{G}_0}({\mathbf{T}}_a^{k - 1}) \\
		& \qquad \qquad +  \alpha\mathbf{T}_a^{k-1} + \rho ( \mathbf{F}_{\mathrm{A},a}^{k-1}\mathbf{F}_{\mathrm{D},a}^{k-1} - {\bm \Omega}_a^{k-1} ) \big) \Big\|_F^2 \\
		\mathrm{s.t.} \; & \left\| \mathbf{T}_a \right\|_F^2 = E, 
	\end{aligned}
	\label{eq:4-25}%
\end{equation}
The following theorem provides the solution to problem \eqref{eq:4-25}.
\begin{theorem}\label{the:1}
	Problem \eqref{eq:4-25} is a quadratically constrained quadratic program (QCQP) with one constraint (QCQP-1), whose closed-form solution can be given by
	\begin{equation}
		\mathbf{T}_a^k = { \sqrt{E}\widetilde{\mathbf{T}}_a^{k-1} } / { \| \widetilde{\mathbf{T}}_a^{k-1} \|_F }.
		\label{eq:4-28}
	\end{equation}
	where $\widetilde{\mathbf{T}}_a^{k-1}$ is defined in SM Appendix \ref{adx:2}.
\end{theorem}
\begin{IEEEproof}
	Please refer to SM Appendix \ref{adx:2}.
\end{IEEEproof}

\subsubsection{Update $\{\mathbf{U}_a , \zeta_a\}$}
Given other variables, the sub-problem of updating $(\mathbf{U}_a , \Psi_a)$ is equivalently rewritten as
\begin{subequations}
	\begin{align}
		\mathop {\min }\limits_{\mathbf{U}_a} \quad & \left\| \mathbf{U}_a - \mathbf{F}_{\mathrm{A},a}^{k-1} \mathbf{F}_{\mathrm{D},a}^{k-1} + {\bm \Lambda}_a^{k-1} \right\|_F^2 \\
		\mathrm{s.t.} \quad & {\overline{\mathrm{MSE}}}_a \left( \mathbf{U}_a , \mathbf{V}_a ,  \zeta_a \right) \le {\gamma _a} , \label{eq:4-29b}
	\end{align}
	\label{eq:4-29}%
\end{subequations}
which can be equivalently rewritten as
\begin{equation}
	\begin{aligned}
		\mathop {\min }\limits_{\mathbf{u}_a} \;  \left\| \mathbf{u}_a - \mathbf{d}_{a} \right\|_F^2 \;\;
		\mathrm{s.t.} \;  \mathbf{u}_a^H \mathbf{G}_{a} \mathbf{u}_a - 2 \Re\left\{ \mathbf{g}_{a}^H \mathbf{u}_a \right\} \le \tilde{\gamma}_a, 
	\end{aligned}
	\label{eq:4-30}
\end{equation}
where 
$\mathbf{u}_a = \mathsf{Vec}(\mathbf{U}_a)$,
$\mathbf{d}_a = \mathsf{Vec}({\mathbf{F}_{\mathrm{A},a}^{k-1} \mathbf{F}_{\mathrm{D},a}^{k-1} - {\bm \Lambda}_a^{k-1}})$,
$\mathbf{G}_a = ( \mathbf{I}_U \otimes \mathbf{G}_{1,a}^H \mathbf{G}_{1,a} )$,
$\mathbf{g}_a = \mathsf{Vec}( \mathbf{G}_{1,a}^H \mathbf{G}_{2,a} )$, and
$\tilde{\gamma}_a = \gamma_a - \|\mathbf{G}_{2,a}\|_F^2$,
with 
$\mathbf{G}_{a,1} = \mathsf{Diag} ( \sqrt{\mu_{a,1}} , \cdots ,\sqrt{\mu_{a,L}}  ) \mathbf{A}_{\mathrm{all}}^H$,
$\mathbf{G}_{a,2} = {\zeta_a^k} {\mathsf{Diag}}( \sqrt{\mu_{a,1}} , \cdots ,\sqrt{\mu _{a,L}}  ) (\mathbf{V}_a^k)^T$, and 
$\mathbf{A}_{\mathrm{all}} = [ \mathbf{a}_\mathrm{T}(\theta_1), \cdots ,\mathbf{a}_\mathrm{T}(\theta_L) ]$.
The following theorem provides the solution to problem \eqref{eq:4-30}.
\begin{theorem}\label{the:2}
	Problem \eqref{eq:4-30} is a convex QCQP-1, whose closed-form solution is derived by analyzing Karush-Kuhn-Tucker (KKT) conditions.
\end{theorem}
\begin{IEEEproof}
	Please refer to SM Appendix \ref{adx:3}.
\end{IEEEproof}

\subsubsection{Update $\mathbf{Z}_{a}$}
Given other variables, the sub-problem of updating $ \mathbf{Z}_a $ can be equivalently rewritten as
\begin{equation}
	\begin{aligned}
		\mathop {\min }\limits_{ \mathbf{Z}_{a} } \; & \left\| \mathbf{Z}_{a} - \{ \mathbf{F}_{\mathrm{A},a}^{k-1} \mathbf{F}_{\mathrm{D},a}^{k-1} \}^H \mathbf{A}_\mathrm{N} + {\bm \Phi}_{a}^{k-1} \right\|_F^2 \\
		\mathrm{s.t.} \; & \max_{t_a} \left\| \mathbf{z}_{a,t_a} \right\|_F^2 \le \Gamma_a , 
	\end{aligned}
	\label{eq:4-38}%
\end{equation}
Problem \eqref{eq:4-38} can be separated into $T_a$ sub-problem as follows
\begin{equation}
	\begin{aligned}
		\mathop {\min }\limits_{ \mathbf{z}_{a,t_a} } \; & \left\| \mathbf{z}_{a,t_a} - \{ \mathbf{F}_{\mathrm{A},a}^{k-1} \mathbf{F}_{\mathrm{D},a}^{k-1} \}^H \mathbf{a}_\mathrm{T}\left( \theta_t\right) + {\bm \phi}_{a,t_a}^{k-1} \right\|_F^2 \\
		\mathrm{s.t.} \; & \left\| \mathbf{z}_{a,t_a} \right\|_F^2 \le \Gamma_a , 
	\end{aligned}
	\label{eq:4-39}%
\end{equation}
which is also a QCQP-1 whose optimal solution can be obtained by analyzing KKT conditions like problem \eqref{eq:4-30} in SM Appendix \ref{adx:3}.
Since above $T_a$ sub-problems are independent to each other, the update of $\mathbf{z}_{a,t_a}$ can be performed in parallel.

\subsubsection{Update $\left\{{\bf F}_{{\rm A} , a}\right\}$}
Given other variables, the sub-problem of updating ${\bf F}_{{\rm A} , a}$ can be equivalently rewritten as
\begin{equation}
	\begin{aligned}
		\mathop {\min }\limits_{ \mathbf{F}_{\mathrm{A},a} } \; & \frac{\rho}{2}\left\| \mathbf{T}_a^k - \mathbf{F}_{\mathrm{A},a} \mathbf{F}_{\mathrm{D},a}^{k-1} + {\bm \Omega}_a^{k-1} \right\|_F^2 \\
		&  + \frac{\varrho}{2} \left\| \mathbf{U}_a^k - \mathbf{F}_{\mathrm{A},a} \mathbf{F}_{\mathrm{D},a}^{k-1}  + {\bm \Lambda}_a^{k-1} \right\|_F^2 \\
		&  + \frac{\lambda}{2} \left\| \mathbf{Z}_{a}^k - \{ \mathbf{F}_{\mathrm{A},a} \mathbf{F}_{\mathrm{D},a}^{k-1} \}^H \mathbf{A}_\mathrm{N} + {\bm \Phi}_{a}^{k-1} \right\|_F^2  \\
		{\rm s.t.} \; & \left| {{{\left[ \mathbf{F}_{\mathrm{A},a} \right]}_{m,n}}} \right| = 1 ,\forall m,n.
	\end{aligned}
	\label{eq:4-42}%
\end{equation}
The following theorem provides the solution to problem \eqref{eq:4-42}.
\begin{theorem}\label{the:3}
	Problem \eqref{eq:4-42} is a quadratic program (QP) with constant modulus constraint, whose closed-form solution at $\ell$-th inner iteration can be given by
	\vspace{-0.5em}
	\begin{equation}
		\mathbf{F}_{\mathrm{A},a}^{[\ell]} = -\exp \left\{ \jmath \angle[ \Bar{\mathbf{W}}_a^{[\ell - 1]}] \right\} .\vspace{-0.5em}
	\end{equation}
	where $\Bar{\mathbf{W}}_a^{[\ell - 1]}$ is defined in SM Appendix \ref{adx:4}.
\end{theorem}
\begin{IEEEproof}
	Please refer to SM Appendix \ref{adx:4}.
\end{IEEEproof}

\subsubsection{Update $\{ \mathbf{F}_{\mathrm{D} ,a }\}$}
Given other variables, the sub-problem of updating $\mathbf{F}_{\mathrm{D},a}$ can be equivalently rewritten as
\begin{equation}
	\begin{aligned}
		\mathop {\min }\limits_{ \mathbf{F}_{\mathrm{D},a} } \;  & \frac{\rho}{2}\left\| \mathbf{T}_a^k - \mathbf{F}_{\mathrm{A},a}^k \mathbf{F}_{\mathrm{D},a} + {\bm \Omega}_a^{k-1} \right\|_F^2 \\
		&  + \frac{\varrho}{2} \left\| \mathbf{U}_a^k - \mathbf{F}_{\mathrm{A},a}^k \mathbf{F}_{\mathrm{D},a} + {\bm \Lambda}_a^{k-1} \right\|_F^2 \\
		&  + \frac{\lambda}{2} \left\| \mathbf{Z}_{a}^k - \{ \mathbf{F}_{\mathrm{A},a} \mathbf{F}_{\mathrm{D},a}^{k-1} \}^H \mathbf{A}_\mathrm{N} + {\bm \Phi}_{a}^{k-1} \right\|_F^2 
	\end{aligned}
\end{equation}
whose closed-form solution can be calculated as
\begin{equation}
	\mathbf{F}_{{\text{D}},a}^k = \left\{  \left( {\mathbf{F}}_{{\text{A}},a}^k \right)^H \mathbf{M}_{1,a}^{k-1} \mathbf{F}_{{\text{A}},a}^k \right\}^{ - 1} \left( {{\mathbf{F}}_{{\text{A}},a}^k} \right)^H {\mathbf{M}}_{2,a}^{k-1} ,
	\label{eq:4-48}
\end{equation}
with ${\mathbf{M}}_{1,a}^{k-1} =  (\rho+\varrho) \mathbf{I}_{N_{\text{T}}} + \lambda \mathbf{A}_\mathrm{N}^H \mathbf{A}_\mathrm{N}$, and
$\mathbf{M}_{2,a}^{k-1} =  \rho (\mathbf{T}_a^k +  {\bm \Omega}_a^{k - 1} )+ \varrho (\mathbf{U}_a^k +  \mathbf{\Lambda }_a^{k - 1} ) + \lambda \mathbf{A}_\mathrm{N} ( \mathbf{Z}_a^k + {\bm \Phi}_a^{k-1} )^H$.

\subsubsection{Update $\{ \mathbf{r}, {\bm \eta}, \{ \mathbf{V}_a \}, {\bm \zeta} \}$}
The optimal solutions to auxiliary variables $\mathbf{r}$ and ${\bm \eta}$ can be derived by first-order derivatives as\vspace{-1em}
\begin{equation}
	{r_u} =  \frac{{{{\left| [{{\bm \Xi}}]_{u,u} \right|}^2}}}{{\sum\limits_{v \in \mathcal{U} ,v \ne u} {{{\left| [{{\bm \Xi}}]_{u,v} \right|}^2}}  + \sigma _{\mathrm{C},u}^2}} , \;
	{\eta_u} = \frac{{\sqrt {{w_u}\left( {1 + {r_u}} \right)} [{{\bm \Xi}}]_{u,u} }}{{\sum\limits_{v \in \mathcal{U}} {{{\left| [{{\bm \Xi}}]_{a,v} \right|}^2}}  + \sigma _{\mathrm{C},u}^2}} .
	\label{eq:4-23}
\end{equation}

Additionally, the optimal solutions to auxiliary variables $\mathbf{V}_a$ and $\zeta_a$ can be directly derived by first-order derivatives as
\begin{equation}
	\mathbf{v}_{a,l}^k =  \sqrt{{P_a}\left(\theta _l\right)}  { \{ \mathbf{a}_\mathrm{T}^H\left( \theta_l \right) \mathbf{U}_{a}^{k}  \}^H  } / { {{{\left\|  \mathbf{a}_\mathrm{T}^H\left( \theta_l \right) \mathbf{U}_{a}^{k}  \right\|}_F}} } , 
	\label{eq:4-37}
\end{equation}
\vspace{-1em}
\begin{equation}
	{\zeta_a^k} = { {\sum\limits_{l = 1}^L { \mu _l \Re \left\{ \mathbf{a}_\mathrm{T}^H\left(\theta_l\right) \mathbf{U}_{a}^{k} \{\mathbf{v}_{a,l}^{k}\}^* \right\}} } } / { {\sum\limits_{l = 1}^L {{\mu _l}{{\left\| \mathbf{v}_{a,l}^{k} \right\|}_F}} } } .
	\label{eq:4-41}
\end{equation}

\vspace{-2em}
\subsection{Summary}
We summarize the above update procedures in Algorithm \ref{alg:2}, where steps 4-11 are distributively updated in corresponding AP until the convergence condition is reached.

\begin{algorithm}[!t]
	\caption{Distributed HBF design for Co-ISACNets.}
	\label{alg:2}
	\LinesNumbered
	\KwIn{System parameters, $k=0$.}
	\KwOut{$\left\{\mathbf{F}_{\mathrm{A},a}\right\}_{\forall a}$, and $\left\{\mathbf{F}_{\mathrm{D},a}\right\}_{\forall a}$.}
	\While{No Convergence}{
		$ k = k + 1 $\;
		\APDo{}{
			Update $\mathbf{r}^k$, ${\bm \eta}^k$, $\{\mathbf{V}_a^k\}$ and ${\bm \zeta}^k$ by \eqref{eq:4-23}-\eqref{eq:4-41}\;
			Update $\mathbf{T}_a^k$ by closed-form solution \eqref{eq:4-28}\;
			Update $\mathbf{U}_a^k$ by analyzing KKT conditions\;
			Update $\mathbf{Z}_{a}^k$ by analyzing KKT conditions\;
			Update $\mathbf{F}_{\mathrm{A},a}^k$ by Algorithm \ref{alg:1} in SM\;
			Update $\mathbf{F}_{\mathrm{D},a}^k$ by closed-form solution \eqref{eq:4-48}\;
			Update dual variables ${\bm \Omega}_a^k$, ${\bm \Lambda}_a^k$, and ${\bm \Phi}_{a}^k$\;
			Exchange local information ${\bm \Xi}_a$ to other APs\;
		}
	}
\end{algorithm}

\subsubsection{Complexity Analysis}
We discuss the complexity of the proposed Algorithm \ref{alg:2} as detailed below.
Specifically, we take $a$-th AP as example, and the main computational complexity for $a$-th comes from steps 4 to 11.
Updating $\mathbf{r}$ and $\bm \eta$ requires complexities of ${\mathscr{O}}( AN_{\mathrm{U}}^2 N_{\mathrm{T}} )$.
Updating $\zeta_a$ with close-form solution requires complexities of ${\mathscr{O}}( L N_{\mathrm{T}} U )$.
Updating $\mathbf{T}_a$ and $\mathbf{V}_a$ with close-form solution need ${\mathscr{O}}( N_{\mathrm{T}} N_{\mathrm{RF}} U )$ and ${\mathscr{O}}( L N_\mathrm{T} U )$, respectively.
Updating $\mathbf{U}_a$ and $\{ \mathbf{z}_{a,t} \}$ by analyzing KKT conditions needs complexities of ${\mathscr{O}}( N_{\mathrm{T}}^2U \log(n) )$ and ${\mathscr{O}}( TU^2 \log(n) )$, respectively.
Updating $\mathbf{F}_{\mathrm{A},a}$ by Algorithm \ref{alg:1} needs complexities of ${\mathscr{O}}( K_1 N_{\mathrm{T}} {N}_{\mathrm{RF}}^2 )$, where $K_1$ is the number of inner iteration.
Updating $\mathbf{F}_{\mathrm{D},a}$ with close-form solution needs ${\mathscr{O}}( N_{\mathrm{T}} {N}_{\mathrm{RF}}^2 + {N}_{\mathrm{RF}}U^2 )$.
Overall, the complexity of the proposed algorithm in each AP is ${\mathscr{O}}( K_0 ( N_{\mathrm{T}}^2U \log(n) + TU^2 \log(n) + K_1 N_{\mathrm{T}} {N}_{\mathrm{RF}}^2 ) )$, where $K_0$ is the number of outer iteration.

\subsubsection{Practical Implementation}

\begin{figure}
		\vspace{-1em}
	\centering
	\includegraphics[width=0.7\linewidth]{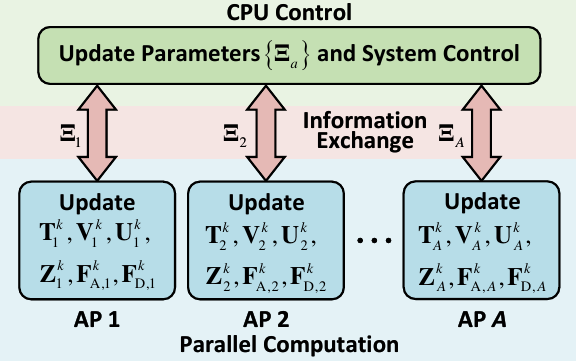}
	\vspace{-0.5em}
	\caption{Workflow diagram of the proposed distributed HBF design algorithm.}
	\vspace{-1em}
	\label{fig:workflow}
	\vspace{-1em}
\end{figure}

The workflow diagram of the proposed distributed algorithm is shown in Fig. \ref{fig:workflow}.
The benefits of implementing the proposed PANDA framework in multi-AP scenarios in practice are summarized as follows. 

\textit{Benefit 1: Enhanced Computational Efficiency.} Based on the proposed PANDA framework, each AP computes $\mathbf{F}_{\mathrm{A},a}$ and $\mathbf{F}_{\mathrm{D},a}$ in parallel to boost the real-time performance and improve the computational efficiency, which is the primary benefit of distributed optimization.

\textit{Benefit 2: Reduced Information Exchange.} Steps 6-12 only require CSI from individual AP without cross message from other APs.
However, update $\mathbf{r}$, $\bm \eta$ and $\mathbf{T}_a$ in steps 4-5 requires the CSI of all APs and auxiliary variables $\mathbf{T}_a, \forall a$ in previous point.
This means each AP must has perfect knowledge of the CSI of all the APs and need frequently change local information $\mathbf{T}_a$.
Fortunately, we find the update of $\mathbf{r}$, $\bm \eta$ and $\mathbf{T}_a$ only requires ${\bm \Xi}_a= \mathbf{H}_a^H \mathbf{T}_a \in \mathbb{C}^{U \times U}$ from other APs.
Therefore, instead of exchanging $\mathbf{T}_a$ and sharing CSI, we can exchange ${\bm \Xi}_a$ as an alternative, which significantly reduces backhaul signaling.

\vspace{-0.75em}
\section{Numerical Results}\label{Sec-5}
\vspace{-0.25em}
In this section, numerical results are presented to evaluate the performance of the proposed PANDA framework in the Co-ISACNet.

\vspace{-1em}
\subsection{Simulation Parameters}
Unless otherwise specified, in all simulations, we assume each AP equipped with $N_{\mathrm{T}} = 32$ transmit antennas and $N_{\mathrm{RF}} = 4$ RF chains serves $U=4$ downlink UEs.
The power of each AP transmitter is set as $E=100\text{ mW}$.
The transmission pulse is $T_0=10$ ms and the bandwidth is $B=150$ MHz.
For the channel model, we adopt Saleh-Valenzuela (SV) channel model \cite{ying2020gmd,samimi201628,wang2020joint} and express $\mathbf{h}_{a,u}$ as $\mathbf{h}_{a,u} = \sqrt{{\hbar_{a,u}}/{N_{\mathrm{P}}}} \big( \kappa_{a,u}^\text{L}\mathbf{a}_\mathrm{T}( \phi_{a,u}^0 ) 
+ \kappa_{a,u}^\text{N}\sum_{p=1}^{N_{\mathrm{p}}-1}\mathbf{a}_\mathrm{T}( \phi_{a,u}^p ) \big)$,
where $N_\mathrm{P} = 10$ denotes the number of paths.
$\phi_{a,u}^0$ denotes the angles of departure (AoD) associated with direct link between the $a$-th AP and $u$-th UE.
$\phi_{a,u}^p$ denotes AoD of $p$-th NLOS path between $a$-th AP and $u$-th UE, which is assumed to follow the uniform distribution.
$\hbar_{a,u} [\text{dB}] = \hbar_0 + 20 \log_{10}( d_{a,u} ) $ denotes the path loss, in which $d_{a,u}$ is the distance between the $a$-th AP and the $u$-th UE and $\hbar_0 = 60$dB is the path loss at the reference distance $d=1$ m.
$\kappa_{a,u}^{\text{L}} = \sqrt{\kappa/(1+\kappa)}$ and $\kappa_{a,u}^{\text{N}}= \sqrt{1/(1+\kappa)}$ are the factor for LoS and NLoS paths, with $\kappa = 6$ being the Rician factor.

Generally, the targets and the clutter sources are located at different spatial angles for different APs.
In the simulation, we calculate the spatial angle $\theta_{o,a}$ of $o$-th target for $a$-th AP through geometrical relation.
Then, the pre-defined spectrum $P_a(\theta_l)$ is given by  $P_a(\theta_l) = 1$ when $\theta_l \in [ \theta_{o,a}-\Delta ,  \theta_{o,a}+\Delta]$, and $P_a(\theta_l) = 0$ otherwise, where $\Delta = 4^\circ$.
The spatial angle $\theta_{q,a}$ of $q$-th clutter source for $a$-th AP can be similarly calculated.
Then, the notch region can be calculated as $\bm{\Theta}_a = [ \theta_{1,a}-{\bar{\Delta}} ,  \theta_{1,a}+{\bar{\Delta}}]  \cup \cdots \cup [ \theta_{Q,a}-{\bar{\Delta}} ,  \theta_{Q,a}+{\bar{\Delta}}]$, where ${\bar{\Delta}} = 2^\circ$.
The radar MSE thresholds and notch depth of different APs are, respectively, set as the same, i.e., $\gamma_a = \gamma$ and $\Gamma_a = \Gamma$.
The radar path loss model is the same as the above communication model.
The radar noise power is set as $\sigma_{\rm{R,u}}^2 = -90$dBm.

\vspace{-1em}
\subsection{Baseline Schemes}
For comparison, the proposed distributed PANDA (Dis-PANDA) algorithm is compared with the following baseline schemes.
\subsubsection{Centralized ADMM (Cen-ADMM)} This scheme solves \eqref{eq:2-7} in a centralized manner, whose detailed procedure can be derived by following framework in the Sec. II-A.
\subsubsection{Semi-distributed two-stage (Semi-Dis TS)} This scheme considers an indirect HBF design method. Specifically, we first design the fully-digital (FD) beamformer for the communication-only case on the CPU side.
Then, we distributively optimize the HBF to approximate to the FD beamformer subject to radar constraints.
\subsubsection{Time-division duplex ISAC with HBF (HBF TDD Mode)} To show more explicitly the advantages of CoCF, we consider a scheme adopting the TDD transmission, where only one AP performs ISAC at the same time.

Besides, the conventional FD architecture is also included to indicate the system performance upper-bound.
Note that all the numerical schemes are analyzed using Matlab 2020b version and performed in a standard PC with Intel(R) CPU(TM) Core i7-10700 2.9 GHz and 16 GB RAM.
The results are averaged over 100 channel realizations.

\vspace{-1em}
\subsection{Scenario 1: Single Target with Clutter Sources}

In this scenario, we evaluate the Co-ISACNet performance in the presence of multiple clutter sources.
Specifically, we assume that there are $A = 3$ APs located at the coordinates (0m, 0m), (90m, 0m), and $(45 \text{m},45\sqrt{3} \text{m})$, respectively.
Additionally, we position the a target at (33m, 26m) and $Q=2$ clutter sources at (28m, 36m) and (51m, 26m), respectively.
The radar MSE is set as $\gamma=4$.

\subsubsection{Impact of the notch depth threshold}
\begin{figure}[!t]
	\centering
	\includegraphics[width = 0.7\linewidth]{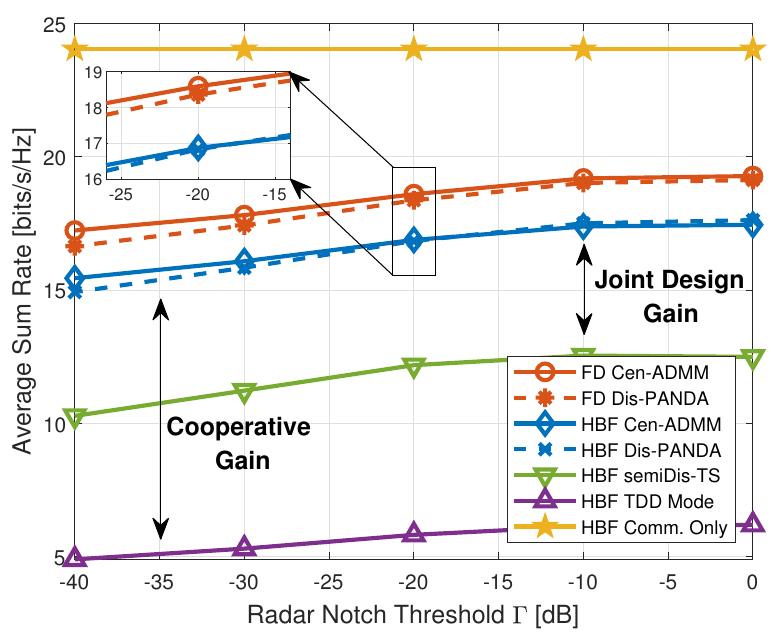}
	\vspace{-1em}
	\caption{The average sum rate versus the intended radar notch requirement $\Gamma$ with $\gamma = 4$.}
	\vspace{-1em}
	\label{fig:Rate_vs_Notch}
\end{figure}
In Fig. \ref{fig:Rate_vs_Notch}, we evaluate the impact of notch depth threshold $\Gamma$ by plotting the average sum rate versus $\Gamma$ when the radar MSE $\gamma=4$.
As expected, there exists a trade-off between communication and radar performance.
Specifically, the achievable sum rate increases with the increase of the radar notch depth threshold $\Gamma$. 
This is because when the intended radar notch depth is smaller, fewer design resources in the optimization problem can be used, resulting degraded communication performance.
The performance of the proposed Dis-PANDA is almost the same as that of the Cen-ADMM regardless of $\Gamma$, where the slight performance loss is due to the fact that PG can only provide a suboptimal solution.
In addition, the proposed Co-ISACNet achieves better performance that the ISAC with TDD mode thanks to the diversity provided by multiple APs.

\subsubsection{Impact of the number of transmit antennas}
\begin{figure}[!t]
	\centering
	\includegraphics[width = 0.7\linewidth]{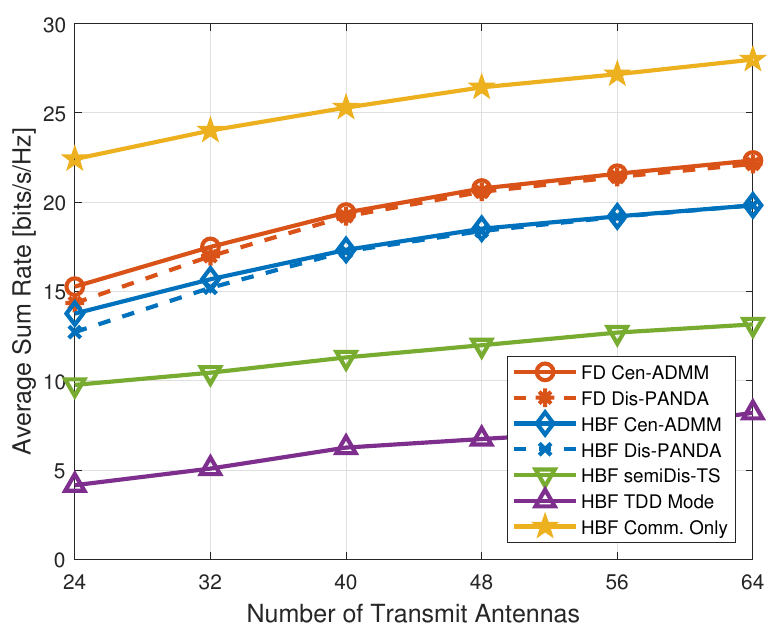}
	\vspace{-1em}
	\caption{The average sum rate versus the number of transmit antennas $N_{\rm TX}$ with $\gamma = 4$ and $\Gamma=-35$dB.}
	\vspace{-1em}
	\label{fig:Rate_vs_NTX}
\end{figure}

In Fig. \ref{fig:Rate_vs_NTX}, we show the sum rate of the proposed Dis-PANDA with respect to the number of transmit antennas $N_{\rm TX}$ when the radar MSE $\gamma=4$ and notch depth $\Gamma=-35$dB.
As expected, the system sum rate will improve as the number of transmit antennas increases, which can offer more antenna diversity and larger beamforming gain.
Besides, we observe that the gap between the proposed joint design method and the indirect Semi-Dis TS design method becomes large, which verifies the essential of joint beamforming design.
Interestingly, we observe that the performance loss between the proposed Dis-PANDA and Cen-ADMM becomes large when $N_\mathrm{TX}$ decreases, which demonstrates the Dis-PANDA can perform well for large-scale MIMO systems.

\subsubsection{Transmit beampattern behaviours}
\begin{figure}[!t]
	\centering
	\subfigure[AP 1]{
		\includegraphics[width = 0.475\linewidth]{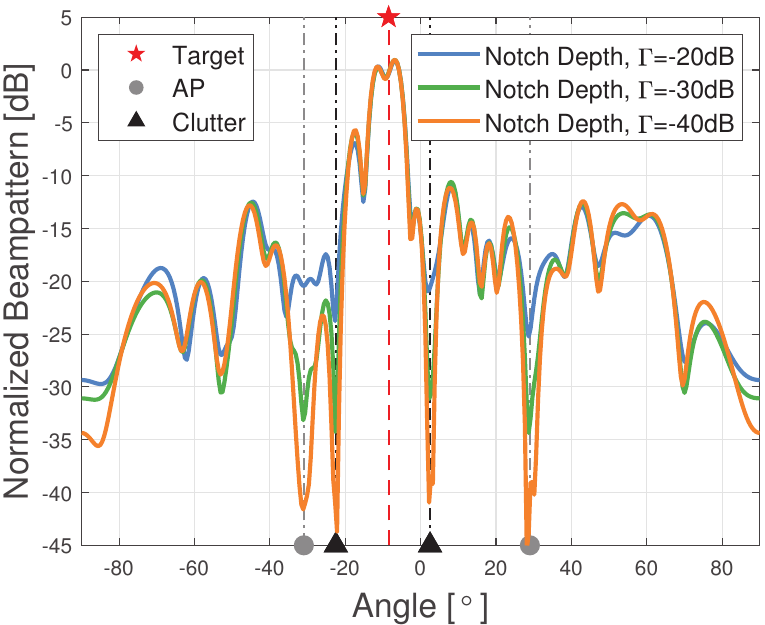}
	}
	\hspace{-1em}
	\subfigure[AP 3]{
		\includegraphics[width = 0.475\linewidth]{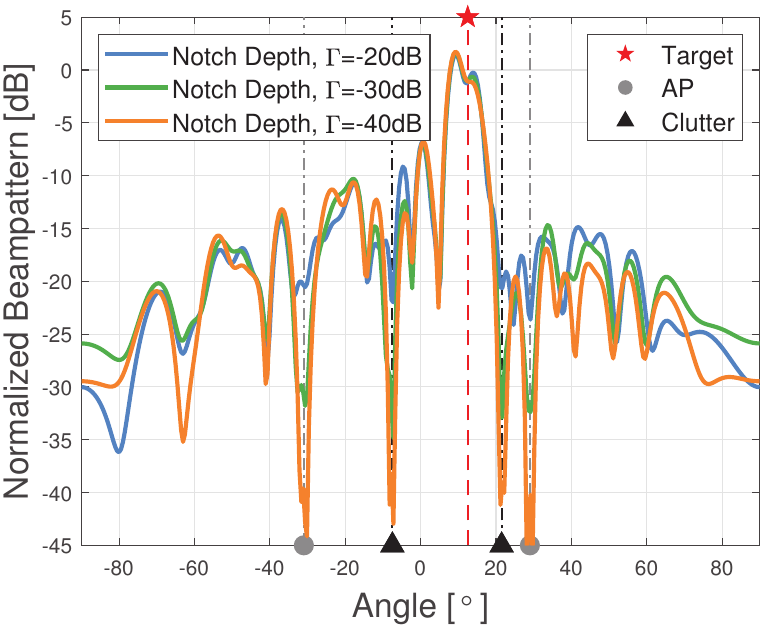}
	}
	\vspace{-1em}
	\caption{The transmit beampattern of AP 1 and AP 3 with $\gamma=4$.}
	\vspace{-1.em}
	\label{fig:Exp2_BP}
\end{figure}

In Fig. \ref{fig:Exp2_BP}, we depict the transmit beampattern of AP 1 and AP 3 with $\gamma=4$.
From the Figs. \ref{fig:Exp2_BP}(a) and \ref{fig:Exp2_BP}(b), we can observe that the transmit power mainly concentrates around the target angle with nearly same mainlobe peak and sidelobe level.
Besides, we also observe that the transmit beampattern can achieve the desired notches at the AP and clutter source angles, which validates the efficiency of the proposed Dis-PANDA algorithm.
Combining Figs. \ref{fig:Exp2_BP} and \ref{fig:Rate_vs_Notch}, we can conclude that a trade-off exists between communication sum rate and radar notch threshold $\Gamma$.
Therefore, we should choose a proper $\Gamma$ to balance communication sum rate and radar beampattern performance.

\subsubsection{Radar detection performance}
\begin{figure}[!t]
	\centering
	\includegraphics[width = 0.7\linewidth]{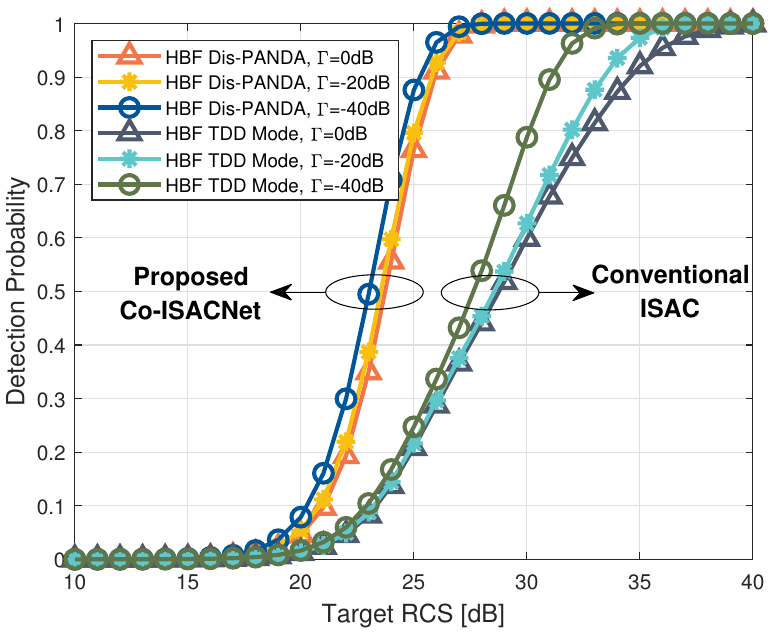}
	\vspace{-1em}
	\caption{The probability of detection $\mathrm{Pr_D}$ versus the target RCS with $\gamma=4$ and false alarm probability $P_{fa}=10^{-5}$.}
	\vspace{-1.em}
	\label{fig:Det_vs_RCS}
\end{figure}
In Fig. \ref{fig:Det_vs_RCS}, we assess the detection performance for the designed HBF considering different notch depth $\Gamma$.
The probability of detection $\mathrm{Pr_D}$ is derived by adopting proposed cooperative sensing detector.
As expected, the smaller $\Gamma$, the higher $\mathrm{Pr_D}$ for all considered algorithms, which is because the smaller $\Gamma$ can achieve sharp nulls at clutter sources angles to resist the strong clutter sources.
Additionally, compared with the conventional ISAC with TDD mode, the proposed Co-ISACNet can achieve a better detection performance benefiting from the diversity provided by multiple APs.

\vspace{-1em}
\subsection{Scenario 2: Multiple Targets}
In this scenario, we consider a scenario where the Co-ISACNet detects multiple targets.
Specifically, we assume that $A = 4$ APs are located at (0m, 0m), (90m, 0m), (0m, 90m) and (90m, 90m), respectively.
Besides, We assume the $o=2$ targets are located in (20m, 50m) and (80m, 45m), respectively.
The notch depth is set as $\Gamma=-30$dB.

\subsubsection{Impact of the radar MSE threshold}
\begin{figure}[!t]
	\centering
	\includegraphics[width = 0.7\linewidth]{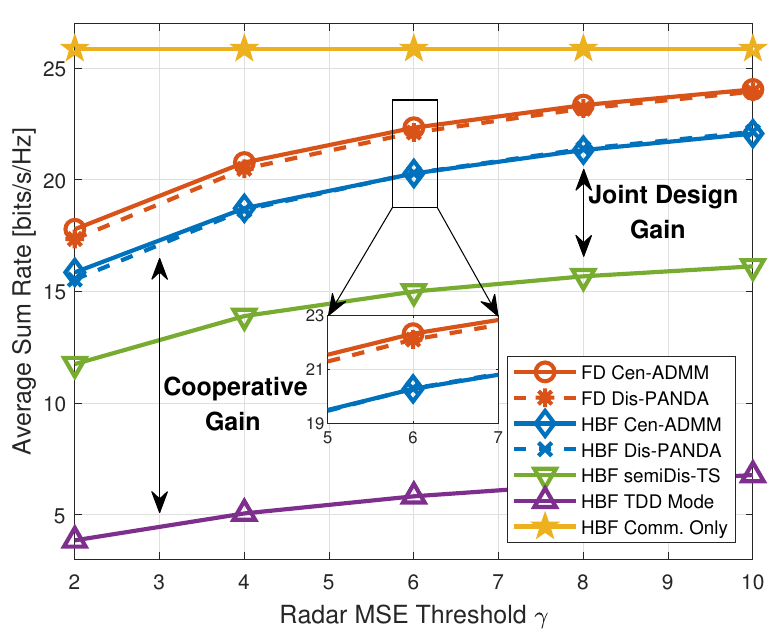}
	\vspace{-1em}
	\caption{The average sum rate versus the radar beampattern MSE requirement $\gamma$ with $\Gamma = -30$ dB.}
	\vspace{-1.5em}
	\label{fig:Rate_vs_Gam}
\end{figure}
In Fig. \ref{fig:Rate_vs_Gam}, we evaluate the impact of radar MSE threshold $\gamma$ by plotting the average sum rate versus $\gamma$ when the notch depth $\Gamma=-30$dB.
A similar conclusion can be drawn from Fig. \ref{fig:Rate_vs_Gam} that the proposed Dis-PANDA achieves nearly the same sum rate as that by Cen-ADMM, and achieves superior performance compared with ISAC with TDD mode and indirect Semi-Dis TS methods.
We can also expect that the sum rate performance becomes better with increasing $\gamma$, which follows the same reason as Fig. \ref{fig:Rate_vs_Notch}.
Besides, we note that compared with radar notch depth $\Gamma$, the radar MSE $\gamma$ has a more significant impact on the sum rate.

\begin{figure}[!t]
	\centering
	\includegraphics[width = 0.7\linewidth]{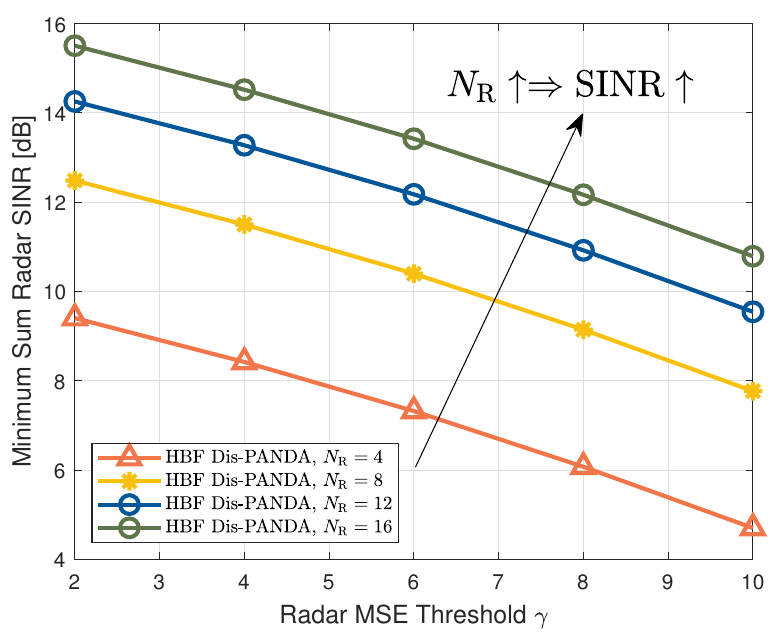}
	\vspace{-1em}
	\caption{The radar SINR versus the radar beampattern MSE requirement $\gamma$ with $\Gamma = -30$ dB.}
	\vspace{-1.5em}
	\label{fig:SINR_vs_MSE}
\end{figure}

In Fig. \ref{fig:SINR_vs_MSE}, we investigate the relationship between radar MSE threshold $\gamma$ and  minimum sum radar SINR
$\min_{o} \{ \sum_{a\in\mathcal{A}}{ \mathrm{SINR}_{a,o} } \} , \forall o$.
In Fig. \ref{fig:SINR_vs_MSE}, we can observe that as the radar MSE threshold $\gamma$ increases, the minimum sum radar SINR decreases. 
This is because a lower radar MSE allows for more concentrated energy around the target, leading to a higher radar SINR.
Furthermore, we find that by increasing the number of radar receive antennas $N_\mathrm{R}$, the minimum sum radar SINR also increases. 
This improvement is attributed to the fact that a larger number of receive antennas can achieve higher receiver beamforming gains.

\subsubsection{Impact of the number of RF chains}
\begin{figure}[!t]
	\centering
	\includegraphics[width = 0.7\linewidth]{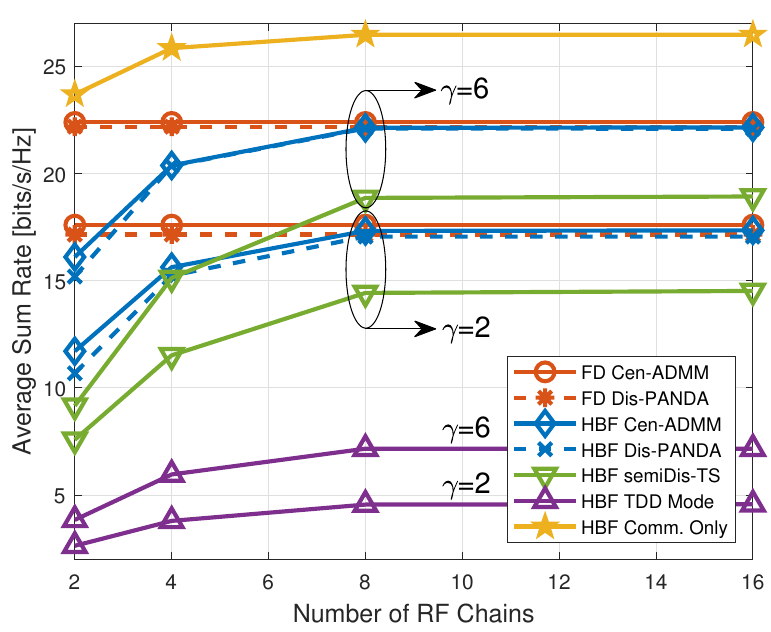}
	\vspace{-1em}
	\caption{The average sum rate versus the number of RF chain $N_{\mathrm{RF}}$.}
	\vspace{-1em}
	\label{fig:Rate_vs_Nrf}
\end{figure}

In Fig. \ref{fig:Rate_vs_Nrf}, we show the average sum rate as a function of the number of RF chains $N_{\mathrm{RF}}$.
As we can predict, for different radar MSE threshold $\gamma$, the average sum rate will first increase and then gradually saturate with the growth of $N_{\mathrm{RF}}$.
In this case, there is little growth in sum rate beyond about $N_\mathrm{RF} \ge 2U$, where the sum rate obtained by the HBF architecture can approximate the that of FD architecture.
Besides, we observe only a slight performance loss when $N_\mathrm{RF} = U$, which confirms that the HBF scheme can dramatically reduce the number of RF chains with acceptable performance loss.

\subsubsection{Transmit beampattern behaviours}
\begin{figure}[!t]
	\centering
	\subfigure[AP 1]{
		\includegraphics[width = 0.472\linewidth]{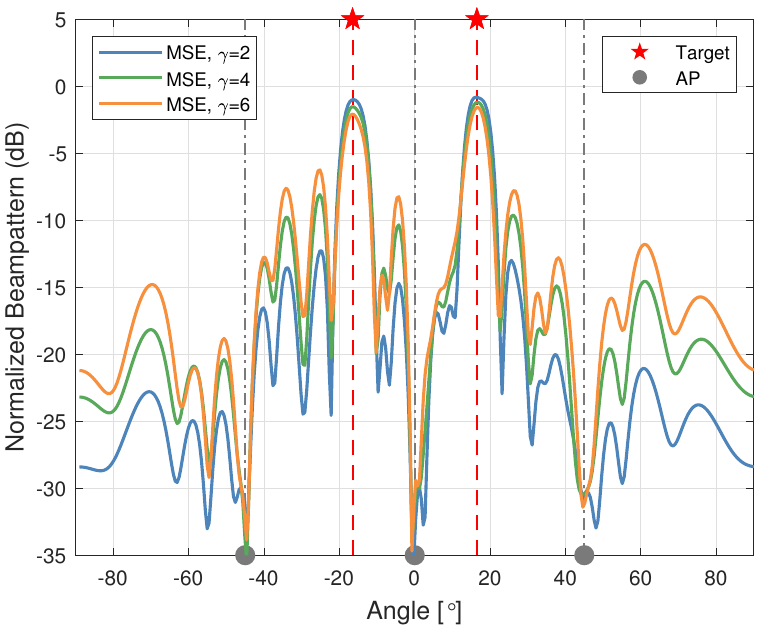}
	}
	\hspace{-1em}
	\subfigure[AP 2]{
		\includegraphics[width = 0.472\linewidth]{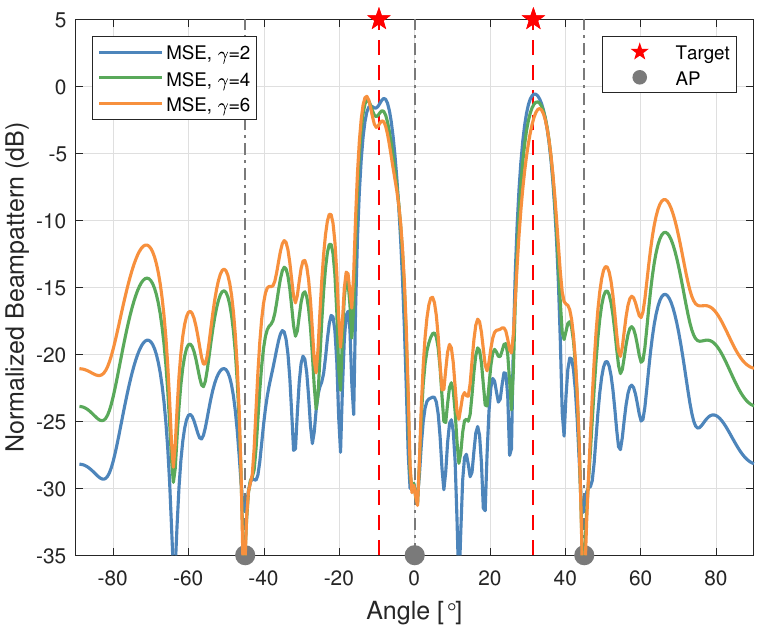}
	}
	\vspace{-1em}
	\caption{The transmit beampattern of AP 1 and AP 2 with $\Gamma=-30$dB.}
	\vspace{-1em}
	\label{fig:Exp1_BP}
\end{figure}

In Fig. \ref{fig:Exp1_BP}, we depict the transmit beampattern of AP 1 and AP 2 with $\Gamma=-30$dB.
The results show that the transmit power mainly concentrates around the two target angles, while achieving notch at the other AP angles.
We also observe that the lower radar MSE $\gamma$, the higher the mainlobe peaks and lower sidelobe level.
Combining Figs. \ref{fig:Exp2_BP} and \ref{fig:Exp1_BP}, we can conclude that the proposed algorithm can flexibly control the beampattern.

\subsubsection{Radar Detection performance}
\begin{figure}[!t]
	\centering
	\includegraphics[width = 0.7\linewidth]{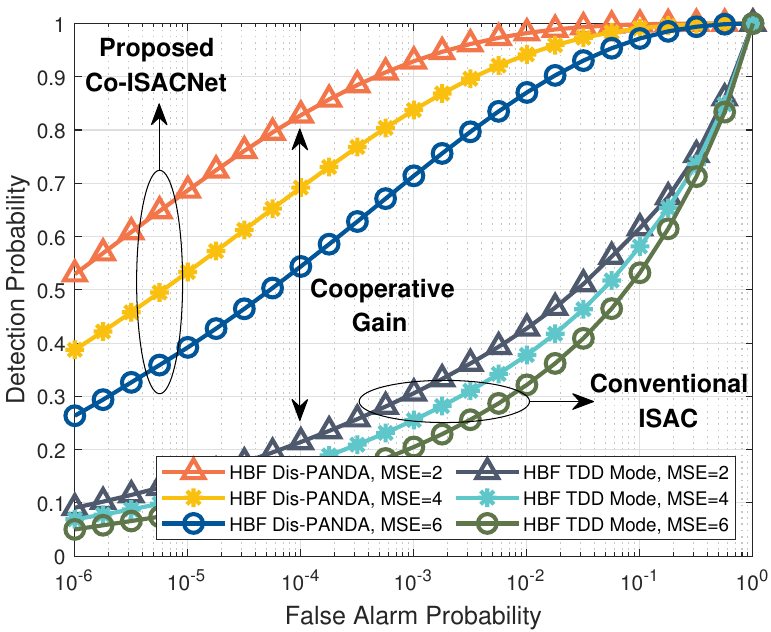}
	\vspace{-1em}
	\caption{The receiver operating characteristic (ROC) curve of the proposed Co-ISACNet.}
	\vspace{-1em}
	\label{fig:PD_vs_Pfa}
\end{figure}

In Fig. \ref{fig:PD_vs_Pfa}, we analyze the radar detection performance of the proposed Co-ISACNet by using the ROC curve.
As can be observed from Fig. \ref{fig:PD_vs_Pfa}, the probability of detection of all methods increases with the probability of false alarm.
Besides, Fig. \ref{fig:PD_vs_Pfa} shows that the HBF with the smaller $\gamma$ has better detection performance, which implies that the beampattern behaviors impact the detection performance.
Additionally, compared with the conventional ISAC with TDD mode, the proposed Co-ISACNet can achieve a remarkable improvement of detection performance thanks to the diversity provided by multiple APs, which verifies the superiority of the proposed Co-ISACNet.

\vspace{-1em}
\subsection{Convergence of the Algorithm}

\begin{figure}[!t]
	\centering
	\subfigure[Scenario 1, $\Gamma=-35$dB.]{
		\includegraphics[width = 0.475\linewidth]{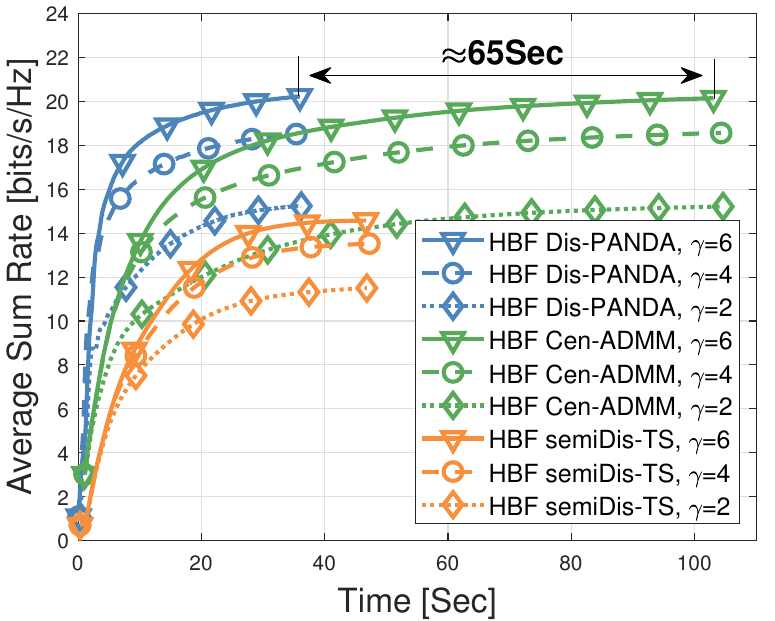}
	}
	\hspace{-1em}
	\subfigure[Scenario 2, $\gamma=4$.]{
		\includegraphics[width = 0.475\linewidth]{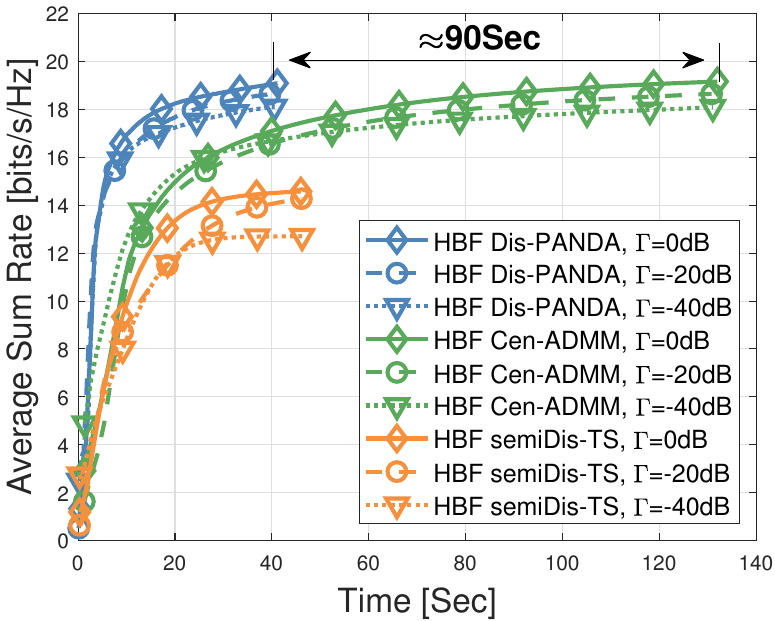}
	}
	\vspace{-1em}
	\caption{The average sum rate versus the CPU time for different scenarios.}
	\vspace{-1em}
	\label{fig:Conv}
\end{figure}

In Fig. \ref{fig:Conv}, we plot the average sum rate versus the CPU time (Sec) for two scenarios.
We can observe from Fig. \ref{fig:Conv} that the proposed Dis-PANDA always converges within finite iterations with different parameter settings.
Besides, the proposed Dis-PANDA achieves almost the same average sum rate as that obtained by Cen-ADMM in less computational time, which validates the time efficiency of the proposed algorithm.
Additionally, the semiDis-TS costs much less computational time but has worse sum rate than Cen-ADMM.

\vspace{-0.5em}
\section{Conclusion}\label{Sec-6}
We proposed a novel Co-ISACNet, where a CPU is deployed to control multiple APs to cooperatively provide communication services to users and detect multiple targets of interest.
A joint HBF optimization problem is formulated with the aim of maximizing the average sum rate of the considered network while satisfying the constraint of beampattern similarity for radar sensing. 
To reduce the computational burden of the CPU and take advantage of the distributed APs, we propose a novel PANDA framework to solve the general joint HBF design problem in a distributed manner.
Then, we customize the proposed PANDA framework to solve the joint HBF design problem of the Co-ISACNet.
Numerical results confirmed that our proposed PANDA based algorithm can achieve nearly the same performance as the centralized algorithm with less computational time.
Moreover, our results revealed that Co-ISACNet is an efficient network to improve both radar and communication performance over conventional ISAC with a single transmitter.

Based on this initial work on Co-ISACNet, there are many issues worth studying for future research, such as asynchronous distributed optimization, wideband waveform design, optimal Co-ISACNet design, and scenarios for target estimation.

\appendices

\vspace{-1em}
\bibliographystyle{IEEEtran}
\bibliography{IEEEabrv,ref_arb.bib}

\newpage
\setcounter{page}{1}
\begin{figure*}
	\begin{center}
		{\Huge Cooperative Integrated Sensing and Communication \\
			\vspace{0.3em}
			Networks: Analysis and Distributed Design}
		
		\vspace{1em}
		{\large Bowen Wang, \textit{Student Member, IEEE}, Hongyu Li, \textit{Member, IEEE}, Fan Liu, \textit{Senior Member, IEEE}, \\ 
			\vspace{0.3em}
			Ziyang Cheng, \textit{Senior Member, IEEE}, and Shanpu Shen, \textit{Senior Member, IEEE}.}
		
		\vspace{1em}
		{\large \textit{(Supplementary Material)}}
	\end{center}
\end{figure*}

\section{ Proof of Cooperative Sensing Detector in \eqref{eq:2-8n}}\label{adx:new-1}
\setcounter{equation}{0}
\renewcommand*{\theequation}{A-\arabic{equation}}

Based on \eqref{eq:2-7n}, we formulate the binary hypothesis test as
\begin{equation}
	\bar{y}_a = \left\{
	\begin{array}{ll}
		\xi_{a,a,0} \bar{x}_{a,a,0} + \sum\limits_{q\in\mathcal{Q}}{ \xi_{a,a,q} \bar{x}_{a,a,q} } + \bar{n}_{\mathrm{R},a}, & \mathcal{H}_1 \\
		\sum\limits_{q\in\mathcal{Q}}{ \xi_{a,a,q} \bar{x}_{a,a,q} } + \bar{n}_{\mathrm{R},a}, & \mathcal{H}_0 \\
	\end{array}
	\right.
\end{equation}

The probability density functions (PDFs) of the received signals $\bar{y}_a$ under hypothesis $\mathcal{H}_0$ and $\mathcal{H}_1$ are respectively given by
\begin{equation}
	\begin{aligned}
		f(\bar{y}_a|\mathcal{H}_{0}) & = \frac{1}{{\pi\sigma_{\mathrm{E},a}^2}}\exp\{-\frac{|\bar{y}_a|^2}{ {\sigma_{\mathrm{E},a}^2} }\} \\
		f(\bar{y}_a|\mathcal{H}_{1}) & = \frac{1}{{\pi\sigma_{\mathrm{E},a}^2}}\exp\{-\frac{|\bar{y}_a - \xi_{a,a,0} \bar{x}_{a,a,0}|^2}{ {\sigma_{\mathrm{E},a}^2} }\} 
	\end{aligned}
\end{equation}
where $\sigma_{\mathrm{E},a}^2 = \bar{\mathbf{x}}_{a}^H\bm{\varsigma}_a\bar{\mathbf{x}}_{a} + \bar{\sigma}_{\rm{R}}^2\|\mathbf{w}_a\|_F^2$,
$\bar{\mathbf{x}}_{a} = [\bar{x}_{a,a,1} ,$ $ \cdots , \bar{x}_{a,a,Q}]^T$ 
and $\bm{\varsigma}_a = \mathsf{Diag}[\varsigma_{a,a,1} , \cdots , \varsigma_{a,a,Q}]^T$.

Then, the joint PDFs of the combined received signal vector $\bar{\mathbf{y}} = [\bar{y}_1 , \cdots, \bar{y}_A]^T$ can be written as
\begin{equation}
	\begin{aligned}
		f(\bar{\mathbf{y}}|\mathcal{H}_0) & = \frac{1}{{\pi^A\prod\limits_{a\in\mathcal{A}} {\sigma_{\mathrm{E},a}^2}}}\exp\{ - \sum\limits_{a\in\mathcal{A}}{ \frac{|\bar{y}_a|^2}{ {\sigma_{\mathrm{E},a}^2} } }\} \\
		f(\bar{\mathbf{y}}|\mathcal{H}_1) & = \frac{1}{{\pi^A\prod\limits_{a\in\mathcal{A}} {\sigma_{\mathrm{E},a}^2}}}\exp\{- \sum\limits_{a\in\mathcal{A}}{\frac{|\bar{y}_a - \xi_{a,a,0} \bar{x}_{a,a,0}|^2}{ {\sigma_{\mathrm{E},a}^2} }}\} 
	\end{aligned}
\end{equation}

To obtain a practical detector, we resort to the GLRT, which is equivalent to replacing all the unknown parameters with their maximum likelihood estimates (MLEs). 
In other words, the GLRT detector in this case is obtained from
\begin{equation}
	\frac{\max_{\bm{\xi}_0}f(\bar{\mathbf{y}}|\mathcal{H}_1)}{f(\bar{\mathbf{y}}|\mathcal{H}_0)} \mathop \gtrless \limits_{\mathcal{H}_0}^{\mathcal{H}_1} \mathcal{T}_0
	\label{eq:pro1-1}
\end{equation}
where $\bm{\xi}_0 = [{\xi}_{1,1,0} , \cdots , {\xi}_{A,A,0}]^T$.
Thus, based on MLE, the estimation of $\bm{\xi}_0$ can be given by
\begin{equation}
	\bar{\xi}_{a,a,0} = \bar{y}_a / \bar{x}_{a,a,0}.
\end{equation}
Inserting this MLE of $\bm{\xi}_0$ into \eqref{eq:pro1-1} leads to
\begin{equation}
	\frac{\max_{\bm{\xi}_0}f(\bar{\mathbf{y}}|\mathcal{H}_1)}{f(\bar{\mathbf{y}}|\mathcal{H}_0)} = \exp\{ \sum\limits_{a\in\mathcal{A}}{ \frac{|\bar{y}_a|^2}{ {\sigma_{\mathrm{E},a}^2} } }\} =  \mathop \gtrless \limits_{\mathcal{H}_0}^{\mathcal{H}_1} \mathcal{T}_0
	\label{eq:pro1-2}
\end{equation}
By simplifying \eqref{eq:pro1-2}, we have 
\begin{equation}
	\sum\limits_{a\in\mathcal{A}}{ \frac{|\bar{y}_a|^2}{ {\sigma_{\mathrm{E},a}^2} } } =  \mathop \gtrless \limits_{\mathcal{H}_0}^{\mathcal{H}_1} \mathcal{T}
\end{equation}

Thereby, this proof is completed.

\section{ Proof of Probability of Detection in \eqref{eq:2-9n}}\label{adx:new-2}
\setcounter{equation}{0}
\renewcommand*{\theequation}{B-\arabic{equation}}

Let $\dot{y}_a = \bar{y}_a / \sigma_{\mathrm{E},a} $. 
The sufficient statistic $\varpi$ in \eqref{eq:2-9n} can be reformulated as $\varpi = \sum_{a\in\mathcal{A}}{ \frac{|\bar{y}_a|^2}{ {\sigma_{\mathrm{E},a}^2} } } = \sum_{a\in\mathcal{A}}{ |\dot{y}_a|^2 }$.
Then we have the following two results:

\begin{enumerate}
	\item Under $\mathcal{H}_0$, the mean and covariance of $\dot{y}_a$ can be expressed as
	\begin{equation}
		\mathbb{E}\{ \dot{y}_a \} = 0, \quad \mathbb{D}\{ \dot{y}_a \} = 1.
		\label{eq:pro2-1}
	\end{equation}
	
	\item Under $\mathcal{H}_1$, the mean and covariance of $\dot{y}_a$ can be expressed as
	\begin{equation}
		\mathbb{E}\{ \dot{y}_a \} = \xi_{a,a,0}\bar{x}_{a,a,0}/\sigma_{\mathrm{E},a}, \quad \mathbb{D}\{ \dot{y}_a \} = 1.
		\label{eq:pro2-2}
	\end{equation}
\end{enumerate}

From the results \eqref{eq:pro2-1} and \eqref{eq:pro2-2} we have the following observations:
\begin{enumerate}
	\item From \eqref{eq:pro2-1}, we know that $\varpi$ under $\mathcal{H}_0$ is standard chi-square distribution with $2A$ DoFs, i.e., $\varpi \sim \chi_{(2A)}^2$. 
	Therefore, the probability of false-alarm can be expressed as
	\begin{equation}
		\begin{aligned}
			\mathrm{Pr_{FA}} & = \mathrm{Pr}(g \ge \mathcal{T}|\mathcal{H}_0) = \mathrm{Pr}(\frac{1}{2}\chi_{(2A)}^2 \ge \mathcal{T}) \\
			& = \mathrm{Pr}(\chi_{(2A)}^2 \ge 2 \mathcal{T}) = 1 - F_{\chi_{(2A)}^2}(1-2 \mathcal{T}).
		\end{aligned}
	\end{equation}
	Thus, the detection threshold can be set according to a desired probability of false-alarm $\mathrm{Pr_{FA}}$, i.e., $\mathcal{T} = \frac{1}{2} F_{\chi_{(2A)}^2}^{-1}( 1-\mathrm{Pr_{FA}} )$, where $F_{\chi_{(A)}^{2}}^{-1}$ represents the inverse CDF of the chi-square distribution with order $A$.
	
	\item From \eqref{eq:pro2-2}, we note that $\varpi$ under $\mathcal{H}_1$ is noncentral chi-square distribution with $2A$ DoFs and noncentrality parameter $\mathfrak{N} = \sum_{a\in\mathcal{A}}{ \frac{|\xi_{a,a,0}\bar{x}_{a,a,0}|^2}{\sigma_{\mathrm{E},a}^2} }$, i.e., $\varpi \sim \chi_{(2A)}^{'2}(\mathfrak{N})$. Thus, the probability of detection can be derived as
	\begin{equation}
		\mathrm{Pr_D}  = \mathbb{Q}_M^A\left( \sqrt{2\mathfrak{N}} , \sqrt{2\mathcal{T}}  \right) 
		\label{eq:pro2-3}
	\end{equation}
\end{enumerate}

By defining $\mathrm{SINR}_a = {|\xi_{a,a,0}\bar{x}_{a,a,0}|^2} / {\sigma_{\mathrm{E},a}^2}$ and plugging  $\mathcal{T} = \frac{1}{2} F_{\chi_{(2A)}^2}^{-1}( 1-\mathrm{Pr_{FA}} )$ into \eqref{eq:pro2-3}, we have
\begin{equation}
	\mathrm{Pr_D} = \mathbb{Q}_M^A\left( \sqrt{2\sum\limits_{a\in\mathcal{A}}{\mathrm{SINR}_a}} , \sqrt{F_{\chi_{(2A)}^{2}}^{-1}(1-\mathrm{Pr_{FA}})}  \right)
\end{equation}
Thus, we complete this proof.

\section{ Proof of Lemma \ref{lem:2} }\label{adx:1}
\setcounter{equation}{0}
\renewcommand*{\theequation}{C-\arabic{equation}}

We start by proving the first property of the proposed PANDA in  Lemma \ref{lem:2} as follows.
For notation simplicity, we define
\begin{equation}
	\begin{aligned}
		& \mathcal{S}_a\left( \mathbf{T}_a , \mathbf{F}_{\mathrm{A},a} , \mathbf{F}_{\mathrm{D},a} , \mathbf{D}_a ; \mathbf{T}_a^{k} \right) \\
		& \qquad = \mathrm{Prox}_{\mathcal{G}_a , \beta}^{\mathcal{X}_a} \left[ \frac{1}{\beta}  \nabla \widetilde{\mathcal{L}}_a \left( {\mathbf{T}}_a , {\mathbf{F}}_{{\text{A}},a} , {\mathbf{F}}_{{\text{D}},a} \right) \right] .
	\end{aligned}
\end{equation} 
Then, according to the procedure of PG, we have the following inequalities
\begin{subequations}
	\begin{align}
		& \sum_{a \in \mathcal{A}}\mathcal{S}_a { \left( \mathbf{T}_a , \mathbf{F}_{\mathrm{A},a}^k , \mathbf{F}_{\mathrm{D},a}^k , \mathbf{D}_a^k ; \mathbf{T}_a^{k} \right) } \notag \\
		& \qquad \qquad \ge \mathcal{L} \left( \{  \mathbf{F}_{\mathrm{A},a}^k \} , \{ \mathbf{F}_{\mathrm{D},a}^k \}, \mathbf{T} , \{ \mathbf{D}_a^k \} \right) , \\
		& \sum_{a \in \mathcal{A}}\mathcal{S}_a { \left( \mathbf{T}_a^k , \mathbf{F}_{\mathrm{A},a}^k , \mathbf{F}_{\mathrm{D},a}^k , \mathbf{D}_a^k  ; \mathbf{T}_a^{k} \right) } \notag \\
		& \qquad \qquad = \mathcal{L} \left( \{  \mathbf{F}_{\mathrm{A},a}^k \} , \{ \mathbf{F}_{\mathrm{D},a}^k \}, \mathbf{T}^k , \{ \mathbf{D}_a^k \} \right) .
	\end{align}
	\label{eq:a-1}%
\end{subequations}
From the minimization of $\mathbf{T}_a , \forall a$ in P1, we have
\begin{equation}
	\begin{aligned}
		& \mathcal{S}_a\left( \mathbf{T}_a^{k+1} , \mathbf{F}_{\mathrm{A},a}^k , \mathbf{F}_{\mathrm{D},a}^k , \mathbf{D}_a^k ; \mathbf{T}_a^{k} \right) \\
		& \qquad \qquad \le \mathcal{S}_a \left( \mathbf{T}_a^k , \mathbf{F}_{\mathrm{A},a}^k , \mathbf{F}_{\mathrm{D},a}^k , \{ \mathbf{D}_a^k \} ; \mathbf{T}_a^{k} \right) .
	\end{aligned}
	\label{eq:a-2}
\end{equation}
Therefore, using \eqref{eq:a-1} and \eqref{eq:a-2}, we obtain
\begin{equation}
	\begin{aligned}
		& \mathcal{L} \left( \{  \mathbf{F}_{\mathrm{A},a}^{k} \} , \{ \mathbf{F}_{\mathrm{D},a}^k \}, \mathbf{T}^{k+1} , \{ \mathbf{D}_a^k \} \right) \\
		& \qquad \qquad \le \mathcal{L} \left( \{  \mathbf{F}_{\mathrm{A},a}^k \} , \{ \mathbf{F}_{\mathrm{D},a}^k \}, \mathbf{T}^k , \{ \mathbf{D}_a^k \} \right) .
	\end{aligned}
\end{equation}
Together with the update of $\mathbf{F}_{\mathrm{A},a}$ and $\mathbf{F}_{\mathrm{D},a}$, we have 
\begin{equation}
	\begin{aligned}
		& \mathcal{L} \left( \{  \mathbf{F}_{\mathrm{A},a}^{k+1} \} , \{ \mathbf{F}_{\mathrm{D},a}^{k+1} \}, \mathbf{T}^{k+1} , \{ \mathbf{D}_a^k \} \right) \\
		& \qquad \qquad \le \mathcal{L} \left( \{  \mathbf{F}_{\mathrm{A},a}^k \} , \{ \mathbf{F}_{\mathrm{D},a}^k \}, \mathbf{T}^k , \{ \mathbf{D}_a^k \} \right) .
	\end{aligned}
	\label{eq:a-3}
\end{equation}
The above inequality \eqref{eq:a-3} shows that the value of the augmented Lagrangian is decreasing with respect to (w.r.t) the primal variables ${ \{  \mathbf{F}_{\mathrm{A},a} \} , \{ \mathbf{F}_{\mathrm{D},a} \}, \mathbf{T} }$ iteratively.
Besides, the augmented Lagrangian is increasing w.r.t the dual variables $\{ \mathbf{D}_a \}$ since the dual ascent is implemented in P4.

This indicates that PANDA exhibits the same convergence behaviour as that of conventional centralized ADMM, which completes the proof.

Then, we prove the second property of the proposed PANDA in Lemma \ref{lem:2}.
Given that we assume $\lim_{k\to \infty} \mathbf{D}_a^{k+1} - \mathbf{D}_a^k = \mathbf{0}$, and  that we update the dual variables in a dual ascent manner in ${\mathrm{P4}}$, we have
\begin{equation}
	\lim_{k\to\infty} \mathbf{T}_{a}^{k+1} - \mathbf{F}_{\mathrm{A},a}^{k+1}\mathbf{F}_{\mathrm{D},a}^{k+1} = \mathbf{0}, \forall a \in \mathcal{A}.
\end{equation}

As demonstrated in \textit{Property 3} of the optimization problem \eqref{eq:3-7}, the equality constraints \eqref{eq:3-7b} and inequality constraints \eqref{eq:3-7c}, denoted as $\mathcal{X}$, are closed and continuous. 
Recalling the update of $\left\{ \mathbf{T}_a \right\}$ is achieved by PG method, as below
\begin{equation}
	\mathbf{T}_a^k = \mathrm{Prox}_{\mathcal{G}_a , \beta}^{\mathcal{X}_a} \left[ \frac{1}{\beta}  \nabla \widetilde{\mathcal{L}}_a \left( {\mathbf{T}}_a^{k - 1} , {\mathbf{F}}_{{\text{A}},a}^{k - 1} , {\mathbf{F}}_{{\text{D}},a}^{k - 1} \right) \right].
\end{equation}
This implies that $\left\{ \mathbf{T}_a \right\}$ is always projected onto the closed and continuous set $\mathcal{X}$, such that $\left\{ \mathbf{T}_a^k \right\}$ remains always bounded.
Since we consider the HBF design problem, the analog beamformers $\left\{ \mathbf{F}_{\mathrm{A},a}^k \right\}$ are subject to the constant modulus constraint.
Therefore, analog beamformers $\left\{ \mathbf{F}_{\mathrm{A},a}^k \right\}$ are also bounded.
The update of digital beamformers $\left\{ \mathbf{F}_{\mathrm{D},a}^k \right\}$ in $\mathrm{P3}$ is an unconstrained optimization problem, whose closed-form solution can be given by
\begin{equation}
	\mathbf{F}_{{\text{D}},a}^k = \left\{  \left( {\mathbf{F}}_{{\text{A}},a}^k \right)^H  \mathbf{F}_{{\text{A}},a}^k \right\}^{ - 1} \left( {{\mathbf{F}}_{{\text{A}},a}^k} \right)^H ( {\mathbf{T}}_{a}^{k-1} - {\mathbf{D}}_{a}^{k-1} ).
\end{equation}
Since $\left\{ \mathbf{T}_a^k \right\}$ and $\left\{ \mathbf{F}_{\mathrm{A},a}^k \right\}$ are bounded, the $\mathbf{F}_{{\text{D}},a}^k$ is bounded.
Therefore, the sequence $\left\{ \left\{ \mathbf{T}_a^k \right\} , \left\{ \mathbf{F}_{\mathrm{A},a}^k \right\} , \left\{ \mathbf{F}_{\mathrm{D},a}^k \right\}  \right\}$ is bounded.
Hence, there exists a stationary point $\left\{ \left\{ \mathbf{T}_a^\star \right\} , \left\{ \mathbf{F}_{\mathrm{A},a}^\star \right\} , \left\{ \mathbf{F}_{\mathrm{D},a}^\star \right\}  \right\}$ such that
\begin{equation}
	\lim_{k\to\infty} \mathbf{T}_{a}^{k} = \mathbf{T}_{a}^\star,
	\; \lim_{k\to\infty} \mathbf{F}_{\mathrm{A},a}^{k} = \mathbf{F}_{\mathrm{A},a}^\star,
	\; \lim_{k\to\infty} \mathbf{F}_{\mathrm{D},a}^{k} = \mathbf{F}_{\mathrm{D},a}^\star.
\end{equation}

Based on the first part of this proof and the boundness of the AL function $\mathcal{L} \left( \{  \mathbf{F}_{\mathrm{A},a} \} , \{ \mathbf{F}_{\mathrm{D},a} \}, \{\mathbf{T}_a\} , \{ \mathbf{D}_a \} \right)$, we have
\begin{equation}
	\begin{aligned}
		& \lim_{k\to\infty} \mathcal{L} \left( \{  \mathbf{F}_{\mathrm{A},a}^{k} \} , \{ \mathbf{F}_{\mathrm{D},a}^k \}, \{\mathbf{T}_a^{k}\} , \{ \mathbf{D}_a^k \} \right) \\
		& \qquad \qquad = \mathcal{L} \left( \{  \mathbf{F}_{\mathrm{A},a}^\star \} , \{ \mathbf{F}_{\mathrm{D},a}^\star \}, \{\mathbf{T}_a^\star\} , \{ \mathbf{D}_a^\star \} \right) ,
	\end{aligned}
\end{equation}
which implies that
\begin{equation}
	\begin{aligned}
		\lim_{k\to\infty} & \mathbf{T}_{a}^{k} - \mathbf{F}_{\mathrm{A},a}^{k}\mathbf{F}_{\mathrm{D},a}^{k} = \mathbf{T}_{a}^\star - \mathbf{F}_{\mathrm{A},a}^\star\mathbf{F}_{\mathrm{D},a}^\star = \mathbf{0}, \forall a \in \mathcal{A} \\
		\lim_{k\to\infty} & \mathbf{D}_a^{k} = \mathbf{D}_a^\star = \mathbf{0}, \forall a \in \mathcal{A}
	\end{aligned}
	\nonumber
\end{equation}
and that the stationary point $\left\{ \left\{ \mathbf{T}_a^\star \right\} , \left\{ \mathbf{F}_{\mathrm{A},a}^\star \right\} , \left\{ \mathbf{F}_{\mathrm{D},a}^\star \right\}  \right\}$ is an optimal solution.

The proof is completed.

\section{ Proof of Theorem \ref{the:1} }\label{adx:2}
\setcounter{equation}{0}
\renewcommand*{\theequation}{D-\arabic{equation}}

Specifically, by introducing a multiplier $\epsilon_{1,a} \in \mathbb{R}$ for the power constant, we obtain the following Lagrangian function
\begin{equation}
	\begin{aligned}
		\mathcal{L}_{1,a} =  & \epsilon_{1,a} ( \left\| \mathbf{T}_a \right\|_F^2 - E ) + \Re\{ \mathsf{Tr} ( {\mathbf{B}}_2  \mathbf{T}_{a} )\}  \\
		& + \frac{\beta}{2} \Big\| \mathbf{T}_a  -  \frac{1}{\beta} \big( -\nabla {\mathcal{G}_0}({\mathbf{T}}_a^{k - 1})  +  \alpha\mathbf{T}_a^{k-1} \\
		& \qquad \qquad + \rho ( \mathbf{F}_{\mathrm{A},a}^{k-1}\mathbf{F}_{\mathrm{D},a}^{k-1} - {\bm \Omega}_a^{k-1} ) \big) \Big\|_F^2 .
	\end{aligned}
\end{equation}
Setting the gradient of the Lagrangian to zero, we have
\begin{equation}
	\mathbf{T}_a = { \widetilde{\mathbf{T}}_a^{k-1}} / { (\alpha + \rho + 2\epsilon_{1,a}) } ,
	\label{eq:4-27}
\end{equation}
where $\widetilde{\mathbf{T}}_a^{k-1} = -{\mathbf{B}}_2^H - \nabla {\mathcal{G}_0}({\mathbf{T}}_a^{k - 1}) + \alpha \mathbf{T}_a^{k-1}+\rho(\mathbf{F}_{\mathrm{A},a}^{k-1} \mathbf{F}_{\mathrm{D},a}^{k-1} - {\bm \Omega}_a^{k-1} )$.
To determine the value of $\epsilon_{1,a}$, we plug \eqref{eq:4-27} into power budget constraint $\| \mathbf{T}_a \|_F^2 = E$ and obtain the optimal solution $\mathbf{T}_a$ as
\begin{equation}
	\mathbf{T}_a^k = \frac{ \sqrt{E}\widetilde{\mathbf{T}}_a^{k-1} }{ \| \widetilde{\mathbf{T}}_a^{k-1} \|_F }.
\end{equation}

Thus, we complete this proof.

\vspace{-0.5em}
\section{ Proof of Theorem \ref{the:2} }\label{adx:3}
\setcounter{equation}{0}
\renewcommand*{\theequation}{E-\arabic{equation}}

Specifically, the Lagrangian function of problem \eqref{eq:4-29} can be expressed as
\begin{equation}
	\mathcal{L}_{2,a} = \left\| \mathbf{u}_a - \mathbf{d}_a \right\|_F^2 
	+ \epsilon_{2,a} \left( \mathbf{u}_a^H \mathbf{G}_{a} \mathbf{u}_a - 2 \Re\left\{ \mathbf{g}_{a}^H \mathbf{u}_a \right\} - \tilde{\gamma}_a \right) ,
\end{equation}
where $\epsilon_{2,a} \in \mathbb{R}^+$ is a multiplier associated with $\mathbf{u}_a^H \mathbf{G}_{a} \mathbf{u}_a - 2 \Re\left\{ \mathbf{g}_{a}^H \mathbf{u}_a \right\} \le \tilde{\gamma}_a$.
Then the corresponding KKT conditions are given by
\begin{subequations}
	\begin{numcases}{}
		\mathbf{u}_a = \left( \mathbf{I}_{N_\mathrm{T}} + \epsilon_{2,a}\mathbf{G}_{a}\right)^{-1} \left( \mathbf{d}_a + \epsilon_{2,a}\mathbf{g}_{a} \right) \label{eq:4-31a}\\
		\mathbf{u}_a^H \mathbf{G}_{a} \mathbf{u}_a - 2 \Re\left\{ \mathbf{g}_{a}^H \mathbf{u}_a \right\} \le \tilde{\gamma}_a \label{eq:4-31b} \\
		\epsilon_{2,a}\left( \mathbf{u}_a^H \mathbf{G}_{a} \mathbf{u}_a - 2 \Re\left\{ \mathbf{g}_{a}^H \mathbf{u}_a \right\} - \tilde{\gamma}_a \right) = 0 \label{eq:4-31c} \\
		\epsilon_{2,a} \ge 0 \label{eq:4-31d}
	\end{numcases}
\end{subequations}
Accordingly, the optimal solution $\mathbf{u}_a$ to \eqref{eq:4-29} can be determined in the following two cases:

\textit{$\bullet$ Case 1:} For $\epsilon_{2,a} = 0$, the optimal solution of $\mathbf{u}_a$ is given by
$ \mathbf{u}_a = \mathbf{d}_a $, which must satisfy the condition \eqref{eq:4-31b}.

\textit{$\bullet$ Case 2:} For $\epsilon_{2,a} > 0$, from \eqref{eq:4-31b} and \eqref{eq:4-31c}, we have
\begin{equation}
	\mathbf{u}_a^H \mathbf{G}_{a} \mathbf{u}_a - 2 \Re\left\{ \mathbf{g}_{a}^H \mathbf{u}_a \right\} = \tilde{\gamma}_a.
	\label{eq:4-32}
\end{equation}
To determine $\epsilon_{2,a}$, we plug \eqref{eq:4-31a} into the equality constraint in \eqref{eq:4-32} and obtain the following equality
\begin{equation}
	\begin{aligned}
		h\left( \epsilon_{2,a} \right) = & \sum_{n=1}^{N_\mathrm{T}U}{ \left| \frac{[\mathbf{d}_a]_n + \epsilon_{2,a}[\mathbf{g}_a]_n }{1 + \epsilon_{2,a}\mu_n }\right|^2 } \\
		& + 2 \Re \left\{ \sum_{n=1}^{N_\mathrm{T}U}{ [\mathbf{g}_a]_n^* \frac{[\mathbf{d}_a]_n + \epsilon_{2,a}[\mathbf{g}_a]_n }{1 + \epsilon_{2,a}\mu_n } } \right\} - \tilde{\gamma}_a \\
		= & 0. 
	\end{aligned}
	\label{eq:4-34}
\end{equation}
where $\mu_n$ is $n$-th singular value of $\mathbf{G}_a$ with $\mu_1 \le \mu_2 \le , \cdots, \le \mu_{N_{\mathrm{T}}U}$.
Then, the derivative of $h( \epsilon_{2,a} )$ is given by 
\begin{equation}
	h'\left( \epsilon_{2,a} \right) = - 2 \sum_{n=1}^{N_\mathrm{T}U}{ \frac{| [\mathbf{g}_a]_n + \epsilon_{2,a}[\mathbf{d}_a]_n |^2}{( 1 + \epsilon_{2,a}\mu_n )^2 } } < 0,
	\label{eq:4-35}
\end{equation}
for all $\epsilon_{2,a} > - 1 / \mu_1$.
Combining \eqref{eq:4-34} and \eqref{eq:4-35}, we know $h(\epsilon_{2,a})$ is monotonic in the possible region  $(0 , +\infty]$, $h(0) > 0$, and $\lim_{\epsilon_{2,a} \to +\infty} h(\epsilon_{2,a}) \le 0$.
Therefore, we can find the unique (optimal) solution $\epsilon_{2,a}^\star$ by bisection or Newton’s method.
By substituting $\epsilon_{2,a}^\star$ into \eqref{eq:4-31a}, the optimal solution to $\mathbf{u}_a$ is obtained.

Thereby, this proof is completed.

\section{ Proof of Theorem \ref{the:3} }\label{adx:4}
\setcounter{equation}{0}
\renewcommand*{\theequation}{F-\arabic{equation}}

Given other variables, the sub-problem of updating ${\bf F}_{{\rm A} , a}$ can be equivalently rewritten as
\begin{equation}
	\begin{aligned}
		\mathop {\min }\limits_{ \mathbf{F}_{\mathrm{A},a} } \; & \frac{\rho}{2}\left\| \mathbf{T}_a^k - \mathbf{F}_{\mathrm{A},a} \mathbf{F}_{\mathrm{D},a}^{k-1} + {\bm \Omega}_a^{k-1} \right\|_F^2 \\
		&  + \frac{\varrho}{2} \left\| \mathbf{U}_a^k - \mathbf{F}_{\mathrm{A},a} \mathbf{F}_{\mathrm{D},a}^{k-1}  + {\bm \Lambda}_a^{k-1} \right\|_F^2 \\
		&  + \frac{\lambda}{2} \left\| \mathbf{Z}_{a}^k - \{ \mathbf{F}_{\mathrm{A},a} \mathbf{F}_{\mathrm{D},a}^{k-1} \}^H \mathbf{A}_\mathrm{N} + {\bm \Phi}_{a}^{k-1} \right\|_F^2  \\
		{\rm s.t.} \; & \left| {{{\left[ \mathbf{F}_{\mathrm{A},a} \right]}_{m,n}}} \right| = 1 ,\forall m,n.
	\end{aligned}
	\label{eq:4-42AA}%
\end{equation}

By defining $\mathbf{W}_{a,1}^{k-1} = [
\sqrt{\frac{\rho}{2}}\mathbf{I}_{N_{\mathrm{T}}} , 
\sqrt{\frac{\varrho}{2}}\mathbf{I}_{N_{\mathrm{T}}} ,
\sqrt{\frac{\lambda}{2}}{\mathbf{A}}_\mathrm{N} ]^H $, and
$\mathbf{W}_{a,2}^{k-1} = [
\sqrt{\frac{\rho}{2}}( \mathbf{T}_a^k + {\bm \Omega}_a^{k-1}), 
\sqrt{\frac{\varrho}{2}}( \mathbf{U}_a^k + {\bm \Lambda}_a^{k-1}),
\sqrt{\frac{\lambda}{2}}( \mathbf{Z}_a^k + {\bm \Phi}_a^{k-1})^H ]^H$,
problem \eqref{eq:4-42AA} can be equivalently rewritten as
\begin{equation}
	\begin{aligned}
		\mathop {\min }\limits_{ \mathbf{F}_{\mathrm{A},a} } \; & f\left( \mathbf{F}_{\mathrm{A},a} \right) = \left\| \mathbf{W}_{a,1}^{k-1}\mathbf{F}_{\mathrm{A},a}\mathbf{F}_{\mathrm{D},a}^{k-1} - \mathbf{W}_{a,2}^{k-1}\right\|_F^2\\
		{\rm s.t.} \; & \left| {{{\left[ \mathbf{F}_{\mathrm{A},a} \right]}_{m,n}}} \right| = 1 ,\forall m,n.
	\end{aligned}
	\label{eq:4-43}%
\end{equation}
To solve the constant modulus constrained quadratic problem \eqref{eq:4-43}, we adopt the block successive upper-bound minimization (BSUM) method.
Specifically, by applying \textit{Lemma 2} again, the upper-bound function of $f( \mathbf{F}_{\mathrm{A},a} )$ at [$\ell$-1]-th inner iteration can be derived as
\begin{equation}
	\begin{aligned}
		& f\left( \mathbf{F}_{\mathrm{A},a} \right)  \le f\left( \mathbf{F}_{\mathrm{A},a}^{[\ell - 1]} \right) + \frac{\tilde{\alpha}}{2} \left\| \mathbf{F}_{\mathrm{A},a} - \mathbf{F}_{\mathrm{A},a}^{[\ell - 1]}\right\|_F^2 \\
		& \quad + \Re\left( \mathsf{Tr} \left\{ ( \nabla f( \mathbf{F}_{\mathrm{A},a}^{[\ell - 1]} ))^H ( \mathbf{F}_{\mathrm{A},a} - \mathbf{F}_{\mathrm{A},a}^{[\ell - 1]}  )\right\}\right) \\
		& = \Re \left( \mathsf{Tr} \left\{  \mathbf{F}_{\mathrm{A},a}^H \bar{\mathbf{W}}_a^{[\ell - 1]} \right\} \right) + \mathrm{c}_3.
	\end{aligned}
\end{equation}
where $\Bar{\mathbf{W}}_a^{[\ell - 1]} = \nabla f(\mathbf{F}_{A,a}^{[\ell - 1]}) - \tilde{\alpha} \mathbf{F}_{A,a}^{[\ell - 1]} $, 
$\nabla f(\mathbf{F}_{A,a}^{[\ell - 1]}) = (\mathbf{W}_{a,1}^{k-1})^H \mathbf{W}_{a,1}^{k-1} \mathbf{F}_{\mathrm{A},a}^{[\ell - 1]}\mathbf{F}_{\mathrm{A},a}^{k - 1} (\mathbf{F}_{\mathrm{A},a}^{k - 1})^H$, and
$\mathrm{c}_3 = f( \mathbf{F}_{\mathrm{A},a}^{[\ell - 1]} ) - \Re ( \mathsf{Tr} \{ (\nabla f(\mathbf{F}_{A,a}^{[\ell - 1]}))^H \mathbf{F}_{A,a}^{[\ell - 1]} \} ) + \tilde{\alpha}/2 \| \mathbf{F}_{A,a}^{[\ell - 1]} \|_F^2 + \tilde{\alpha} N_{\mathrm{T}}U$.
Then, $\mathbf{F}_{\mathrm{A},a}$ can be updated by iteratively solving the following problem
\begin{equation}
	\begin{aligned}
		\mathop {\min }\limits_{ \mathbf{F}_{\mathrm{A},a} } \;  \Re \left( \mathsf{Tr} \left\{  \mathbf{F}_{\mathrm{A},a}^H \bar{\mathbf{W}}_a^{[\ell - 1]} \right\} \right) \;\;
		{\rm s.t.} \;  \left| {{{\left[ \mathbf{F}_{\mathrm{A},a} \right]}_{m,n}}} \right| = 1 ,\forall m,n,
	\end{aligned}
\end{equation}
whose closed-form solution can be given by 
\begin{equation}
	\mathbf{F}_{\mathrm{A},a}^{[\ell]} = -\exp \left\{ \jmath \angle[ \Bar{\mathbf{W}}_a^{[\ell - 1]}] \right\} .
\end{equation}

The overall algorithm for updating analog beamformer $\mathbf{F}_{\mathrm{A},a}$ is summarized in Algorithm \ref{alg:1}.
\begin{algorithm}[!t]
	\renewcommand{\thealgocf}{F-1}
	\caption{Analog beamformer $\mathbf{F}_{\mathrm{A},a}$ design.}
	\label{alg:1}
	\LinesNumbered
	\KwIn{System parameters,  $\mathbf{F}_{\mathrm{A,a}}^{k-1}$}
	\KwOut{$\mathbf{F}_{\mathrm{A,a}}^{k}$}
	\textbf{Initialization: }$\mathbf{F}_{\mathrm{A},a}^{[0]} = \mathbf{F}_{\mathrm{A,a}}^{k-1} $, $\ell = 1$\;
	\Repeat{Convergence}{
		$\ell = \ell + 1$\;
		Calculate $\Bar{\mathbf{W}}_{a}^{[\ell-1]}$\;
		Update $\mathbf{F}_{\mathrm{A},a}^{[\ell]} = -\exp \{ \jmath \angle[ \Bar{\mathbf{W}}_a^{[\ell-1]}] \}$\;
	}
	\textbf{Return:} $\mathbf{F}_{\mathrm{A,a}}^{k} = \mathbf{F}_{\mathrm{A,a}}^{[\ell]}$\;
\end{algorithm}

\end{document}